\documentclass[prd,twocolumn,floatfix,amsmath,nofootinbib,amssymb,floatfix]{revtex4}
\usepackage{graphicx,color,dcolumn,booktabs,bm}
\usepackage{longtable,lscape}
\usepackage{txfonts}
\usepackage{overpic}
\usepackage{amssymb}
\usepackage{indentfirst}
\usepackage{feynmf}   
\usepackage{slashed}  
\usepackage{cases}
\usepackage{color}
\usepackage{multirow}
\usepackage{epstopdf}
\usepackage{enumerate}
\usepackage{graphicx,color,dcolumn,booktabs,bm}
\usepackage[colorlinks, citecolor=blue,anchorcolor=red,menucolor=red, linkcolor=red,filecolor=red,urlcolor=blue,frenchlinks=red]{hyperref}

\graphicspath{{Figures/}} %

\begin{document}

\title{Identifying the contribution of higher $\rho$ mesons around 2 GeV in the $e^+ e^- \to \omega\pi^0$ and $e^+ e^- \to \rho \eta^{\prime}$ processes}
\author{Qin-Song Zhou$^{1,2}$}\email{zhouqs13@lzu.edu.cn}
\author{Jun-Zhang Wang$^{1,2}$}\email{wangjzh2012@lzu.edu.cn}
\author{Xiang Liu$^{1,2,3,4}$}\email{xiangliu@lzu.edu.cn}
\author{Takayuki Matsuki$^{5}$}\email{matsuki@tokyo-kasei.ac.jp}
\affiliation{$^1$School of Physical Science and Technology, Lanzhou University, Lanzhou 730000, China\\
$^2$Research Center for Hadron and CSR Physics, Lanzhou University $\&$ Institute of Modern Physics of CAS, Lanzhou 730000, China\\
$^3$Lanzhou Center for Theoretical Physics, Key Laboratory of Theoretical Physics of Gansu Province,
and Frontiers Science Center for Rare Isotopes, Lanzhou University, Lanzhou 730000, China\\
$^4$Joint Research Center for Physics, Lanzhou University and Qinghai Normal University, Xining 810000, China\\
$^5$Tokyo Kasei University, 1-18-1 Kaga, Itabashi, Tokyo 173-8602, Japan}

\date{\today}

\begin{abstract}
The properties of the light vector meson states around 2.0 GeV have been poorly understood for a long time, which has become a barrier to the expansion to higher light vector meson spectrum. Recently, the BESIII collaboration released the measurements of the $e^+ e^- \to \omega\pi^0$ and $e^+ e^- \to \rho\eta^{\prime}$ reactions above 2.0 GeV, both of which are ideal processes to study the isovector $\rho$ meson family. In this work, through carrying out a combined analysis of the Born cross section data for the above two processes with the theoretical support on mass spectrum, and production and strong decay behaviors of the $\rho$ meson family around 2.0 GeV, we identify the enhancement structure near 2034 MeV observed in $e^+e^-\to\omega\pi^0$ to be the interference contribution from two resonances $\rho(1900)$ and $\rho(2150)$, and another enhancement structure at 2111 MeV reported in $e^+e^-\to\rho\eta^{\prime}$ to be the contributions from $\rho(2000)$ and $\rho(2150)$. This conclusion means that the $e^+e^-\to\omega\pi^0$ and $e^+e^-\to\rho \eta^{\prime}$ are the excellent golden channels to establish $\rho(1900)$ and $\rho(2000)$, 
especially for a $D$-wave state $\rho(2000)$, whose experimental search in the $e^+e^-$ collision should be quite challenging. The relevant cross section measurements with higher precision are expected in the future BESIII and Belle II experiments.
\end{abstract}

\maketitle

\section{Introduction}\label{section1}
The $e^+ e^-$ annihilation is an ideal platform to produce vector particles like $\rho$, $\omega$, $\phi$, and $J/\psi$.
Recently, the Born cross sections of the $e^+e^- \to \omega \pi^0$ \cite{BESIII:2020xmw} and $e^+e^-\to\eta^{\prime}\pi^+\pi^-$ \cite{BESIII:2020kpr} processes at different center-of-mass energies between 2.0 and 3.08 GeV were measured by the BESIII Collaboration.
One resonance with a mass of $2034\pm13$ MeV and a width of $234\pm30$ MeV was observed in $e^+e^- \to \omega \pi^0$, while another resonance with a mass of $2111\pm43\pm25$ MeV and a width of $135\pm35\pm30$ MeV was observed in $e^+e^- \to \eta^{\prime} \pi^+\pi^-$.
Since the final state of $\pi^+\pi^-$ is found to be dominant in the $\rho(770)$ decay, and the non-$\rho(770)$ contribution is less than $10\%$ by analyzing the Dalitz plots \cite{BESIII:2020kpr}, {the three-body reaction $e^+e^-\to\eta^{\prime}\pi^+\pi^-$ can be treated as $e^+e^-\to\rho\eta^{\prime}\to\eta^{\prime}\pi^+\pi^-$}.
Obviously, the $\omega \pi^0$ and $\rho\eta^{\prime}$ with quantum number $I^G=1^+$ are clean channels to explore the $\rho$ meson family, where the intermediate $\omega$ and $\phi$ states are forbidden.
Thus, the experimental results of these two processes measured by the BESIII Collaboration may provide a good opportunity to study the $\rho$ meson family around 2 GeV.

\begin{figure}
  \centering
  \begin{tabular}{c}
  \includegraphics[width=180pt]{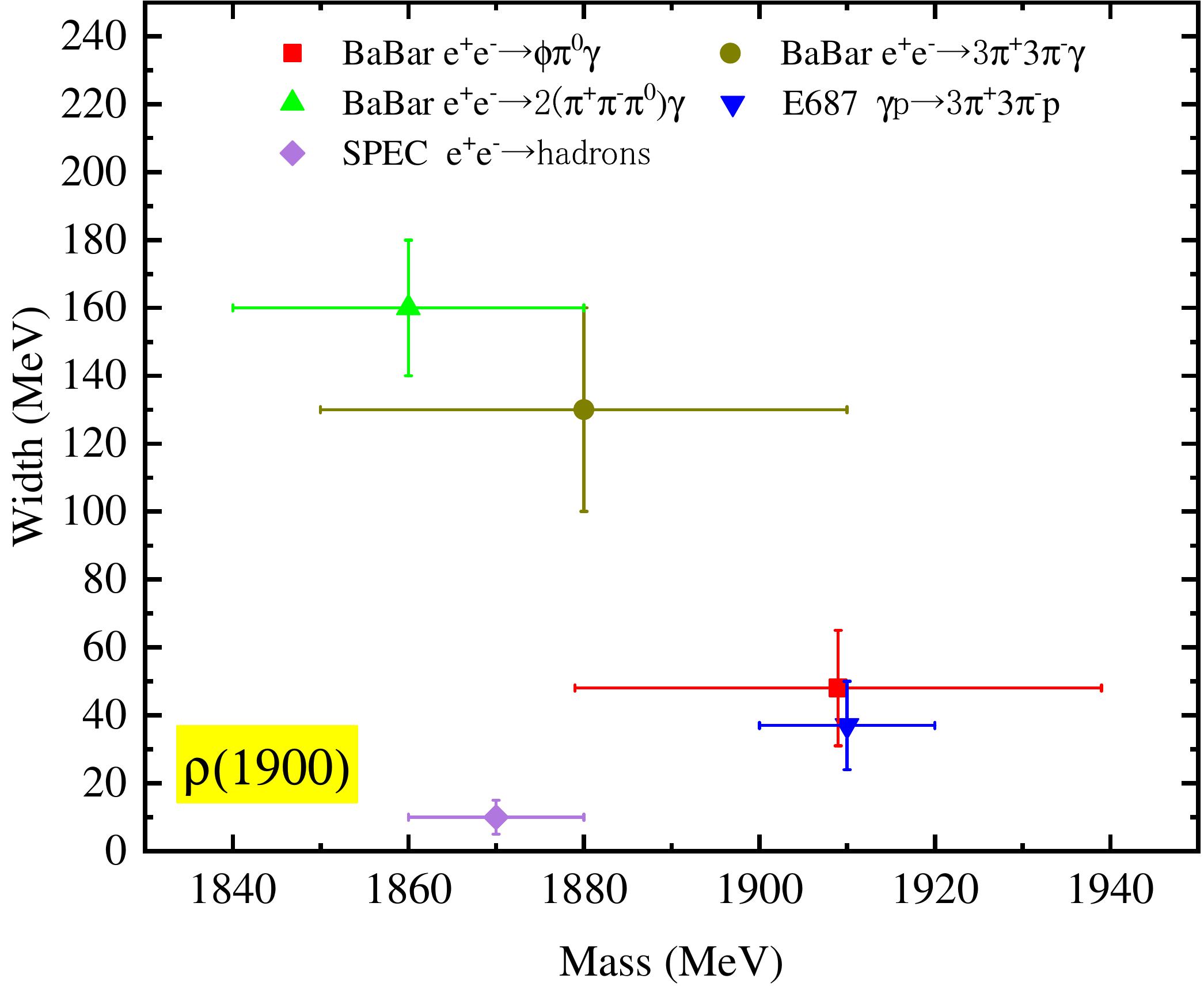}\\
  \includegraphics[width=180pt]{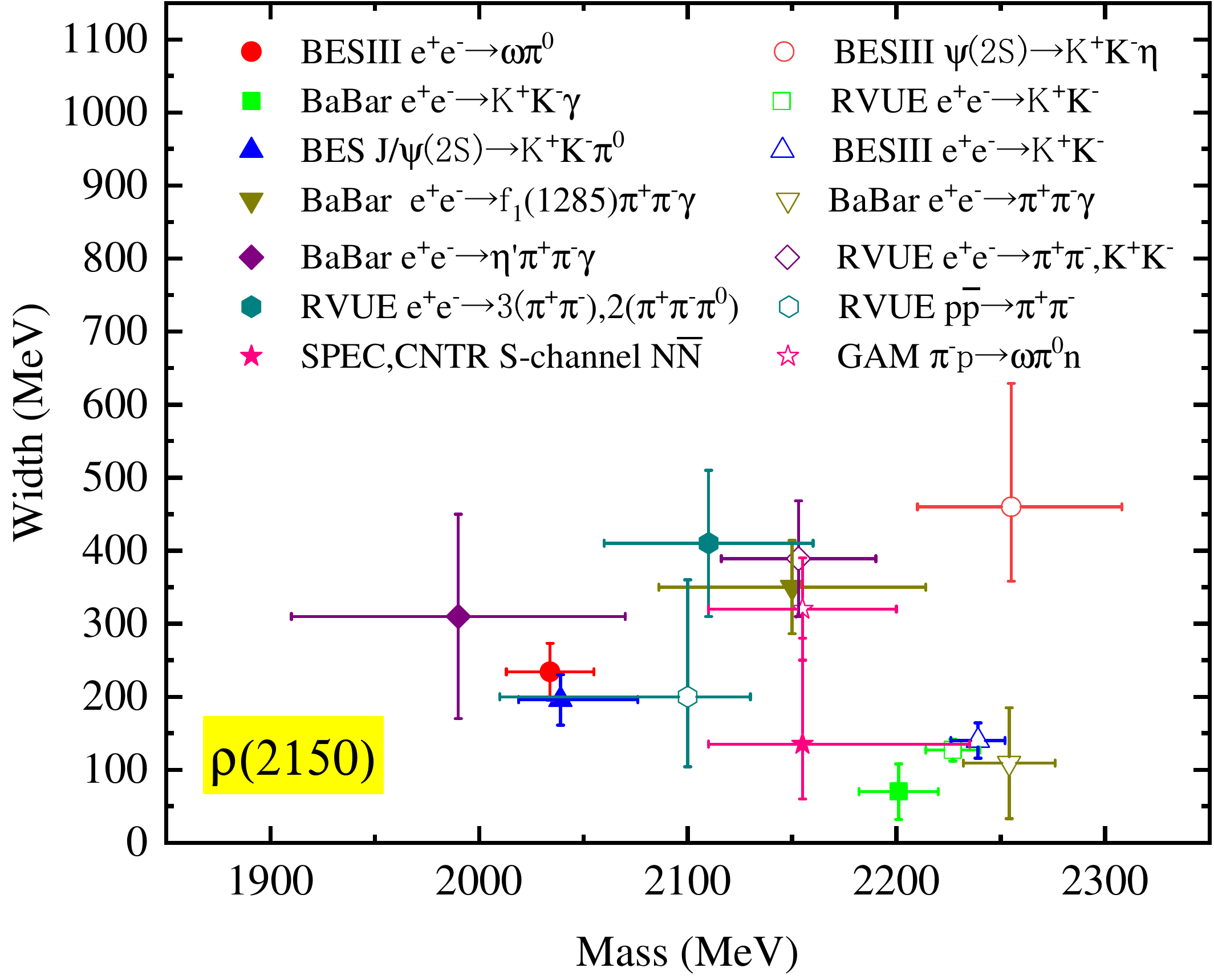}\\
  \end{tabular}
  \caption{The comparison of the resonance parameters of the $\rho(1900)$ and $\rho(2150)$ measured by different experiments collected in PDG \cite{ParticleDataGroup:2020ssz}.}\label{experimentalparameter}
\end{figure}

The $\rho$ meson family around 2 GeV is far from being established.
There are many $\rho$ meson states collected in the Particle Data Group (PDG) \cite{ParticleDataGroup:2020ssz}.
In the mass region around 2 GeV, there are three $\rho$ meson states, which are $\rho(1900)$, $\rho(2000)$, and $\rho(2150)$ .
The $\rho(1900)$  was first observed in the measurement of $e^+e^-\to \rm{hadrons}$ by the FENICE Collaboration \cite{FENICE:1996xlc}, which shows a very narrow dip with a width of 10 MeV near 1.87 GeV.
Thereafter, the $\rho(1900)$ was also reported in $\gamma p\to 3\pi^{+}3\pi^{-}p$ by E687 \cite{Frabetti:2001ah,Frabetti:2003pw} and $e^+e^-\to \phi \pi^0 \gamma$ by BaBar \cite{BaBar:2007ceh} as a narrow state with width of 24 to 65 MeV.
However, the measurements of both  $e^+e^-\to2(\pi^+\pi^-\pi^0)\gamma$ and $e^+e^-\to3\pi^+3\pi^-\gamma$ by BaBar  \cite{BaBar:2006vzy} indicate that the width of $\rho(1900)$ is more than 100 MeV.
In fact, the resonance parameters of the $\rho(1900)$ is not well determined.
Referring to PDG \cite{ParticleDataGroup:2020ssz}, we can find abundant experimental results about $\rho(2150)$.
However, the resonance parameters of the $\rho(2150)$ are very diverse among different experiments.
If considering the experimental uncertainties among different measurements, the mass and width of the $\rho(2150)$ are in the range of 1910-2310 MeV and 32-630 MeV, respectively.
The $\rho(2000)$ has been collected in PDG as the ``{\it further state}" \cite{ParticleDataGroup:2020ssz}. The reason for this is that $\rho(2000)$ was first reported by an amplitude analysis of the data of $p\bar{p}\to\pi\pi$ \cite{Hasan:1994he}, and then its existence was confirmed by a combined analysis for the $e^+e^-\to\omega\eta\pi^0$ and $\omega\pi$ \cite{Bugg:2004xu}.
In Fig. \ref{experimentalparameter}, we show the messy situation of the measurements of the resonance parameters of these $\rho$ meson states around 2 GeV, where we take $\rho(1900)$ and $\rho(2150)$ as an example.
We hope that experimentalists make much more  effort to establish these $\rho$ states around 2 GeV.

Until now, there have been some theoretical studies on these $\rho$ states around 2 GeV. The analysis of mass spectrum of the $\rho$ meson family  has been done in \cite{He:2013ttg,Li:2021qgz,Feng:2021igh,Anisovich:2000kxa,Masjuan:2012gc,Masjuan:2013xta,Bugg:2012yt,Wang:2021gle}, which  suggested that three $\rho$ meson states $\rho(1900)$, $\rho(2000)$, and $\rho(2150)$ are generally assigned to the $\rho(3^3S_1)$, $\rho(2^3D_1)$, and $\rho(4^3S_1)$ states, respectively.
For the observed two enhancement structures reported in $e^+e^-\to \omega\pi^0$ and $e^+e^-\to\rho\eta^{\prime}$, the authors of Ref. \cite{Li:2021qgz}  adopted the modified Godfrey-Isgur model to study the $\rho$ meson spectroscopy, and suggested the newly observed structure in $e^+e^-\to\omega\pi^0$ to be the same state as the $\rho(2000)$ with the $2^3D_1$ quantum number.
In Ref. \cite {Yu:2021ggd}, Yu {\it et al.} estimated the masses of the $D$-wave vector $\rho$ mesons by the QCD sum rule, which also supports the above interpretation. In these theoretical studies, one usually focuses on whether or not the resonance parameters of these observed structures can be reproduced, and then comes to a conclusion of the properties of these $\rho$ states. In fact, besides the measurement of resonance parameters for the reported enhancement structures $e^+e^-\to\omega\pi^0$ \cite{BESIII:2020xmw} and $e^+e^-\to\rho\eta^{\prime}$ \cite{BESIII:2020kpr}, there exist abundant data of the cross sections of $e^+e^-\to\omega\pi^0$ \cite{BESIII:2020xmw} and $e^+e^-\to\rho\eta^{\prime}$ \cite{BESIII:2020kpr}, which should be paid more attention since this information has a close relationship with the mass spectrum and decay behavior of these discussed states and is a crucial step to decode their properties \cite{Wang:2020kte,Wang:2021gle}.

Along this line, in this paper, we perform the study of the cross sections of $e^+e^-\to\omega\pi^0$ and $e^+e^-\to\rho\eta^{\prime}$. Checking the experimental results of the $e^+e^-\to\omega\pi^0$ process, we notice that it is not suitable to simply consider the observed enhancement structure existing in $e^+e^-\to\omega\pi^0$ due to the $\rho(2000)$ contribution. Usually, the dilepton width of the $D$-wave vector meson states should be suppressed compared to the case of the corresponding $S$-wave vector mesons, where the $\rho(2000)$ is a typical $D$-wave state. We  want to ask a natural question: why is the $S$-wave meson state near the $\rho(2000)$ absent in the cross section data? Here, we only take the $e^+e^-\to\omega\pi^0$ process as an example.
Facing these data of the cross sections of $e^+e^-\to\omega\pi^0$ \cite{BESIII:2020xmw} and $e^+e^-\to\rho\eta^{\prime}$ \cite{BESIII:2020kpr}, we should identify the contribution from different $\rho$ meson states to these obtained cross sections, which is a good opportunity to establish the $\rho$ meson states around 2 GeV.

In order to solve these problems pointed above, in this paper,  we perform a combined analysis to the experimental data of Born cross sections of $e^+e^-\to \omega\pi^0$ and $e^+e^-\to \rho\eta^{\prime}$ with the theoretical support from spectroscopy \cite{He:2013ttg,Wang:2020kte}. Subsequently, we show that the enhancement structure near 2034 MeV observed in $e^+e^-\to \omega \pi^0$ cannot be explained as a single $\rho(2000)$ state, where the $\rho(1900)$ and $\rho(2150)$ are the main source of the enhancement structure near 2034 MeV of $e^+e^-\to \omega \pi^0$. Our study indicates that another enhancement structure near 2111 MeV existing in $e^+e^-\to \rho\eta^{\prime}$ can explain it due to the contributions from the $\rho(2000)$ and $\rho(2150)$. With the accumulation of more experimental data, we  believe the above observations made in this work can be further tested.

This paper is organized as follows. After the introduction, we present our theoretical framework of analyzing $e^+e^-\to \omega \pi^0$ and $e^+e^-\to \rho\eta^{\prime}$ by considering the higher $\rho$ mesons as the intermediate states in Sec. \ref{section2}. In Sec. \ref{section3}, we study the experimental data of the Born cross sections of $e^+e^-\to \omega\pi^0$ and $e^+e^-\to \rho\eta^{\prime}$. Finally, this paper ends with a short summary in Sec. \ref{section4}.

\section{$e^+ e^- \to \omega\pi^0$ and $e^+ e^- \to\rho\eta^{\prime}\to\eta^{\prime}\pi^+\pi^-$ processes}\label{section2}

\begin{figure}
  \centering
  \begin{tabular}{cc}
  \includegraphics[width=120pt]{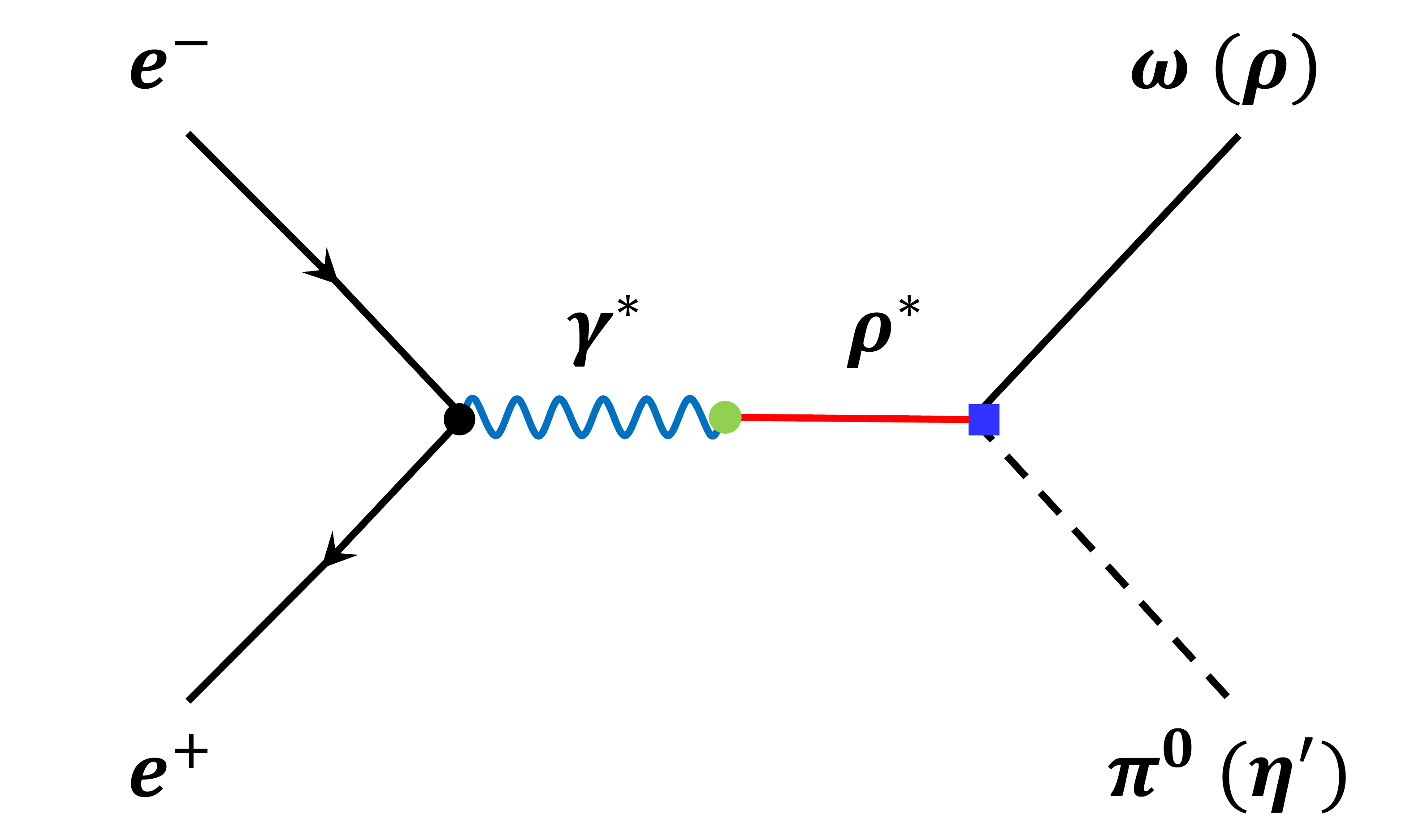}&\includegraphics[width=120pt]{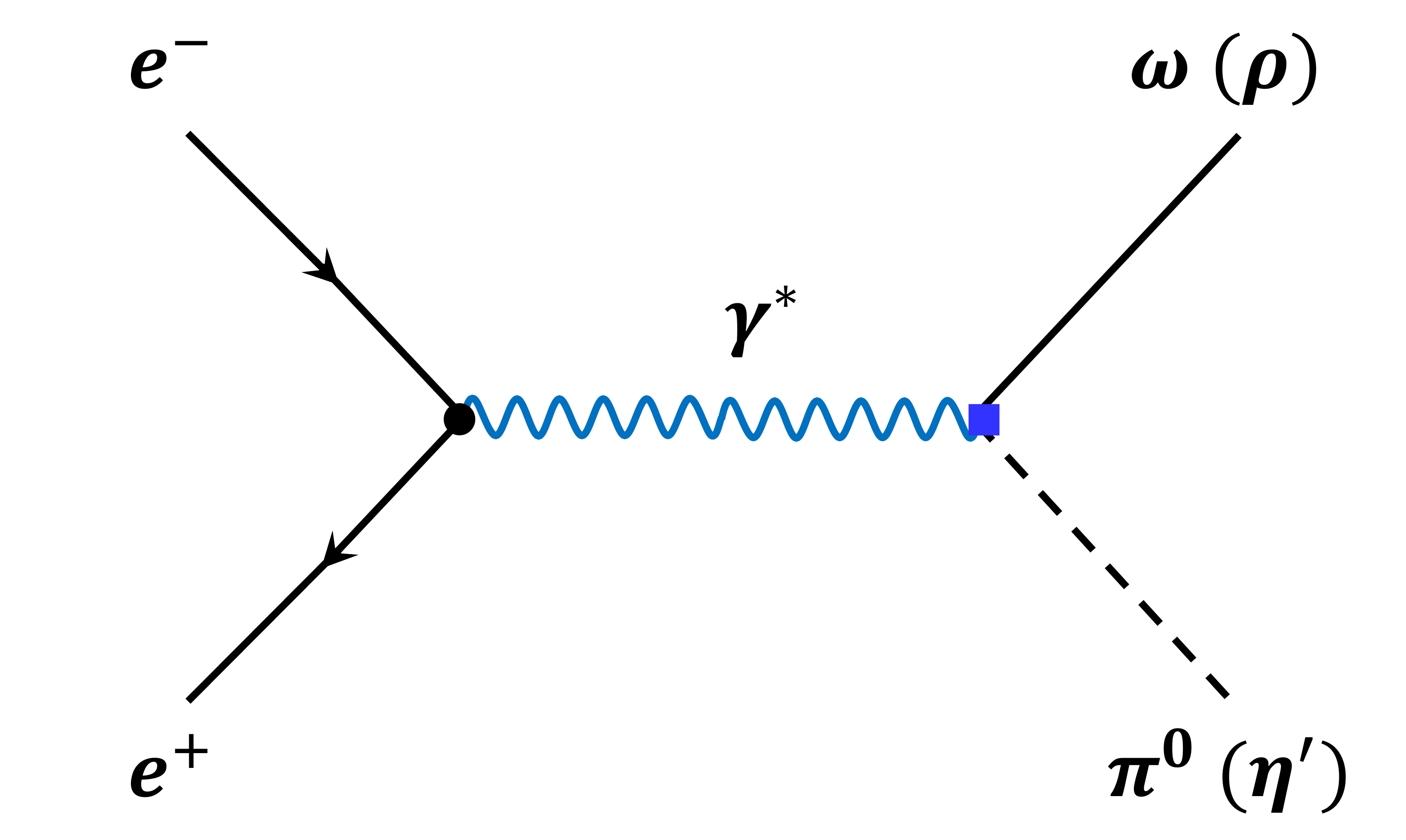}\\
  (a)&(b)\\
  \end{tabular}
  \caption{The Feynman diagrams for depicting the $e^+ e^- \to \omega\pi^0$ and $\rho\eta^\prime$ reactions. Here, diagram (a) is from the contribution of the intermediate $\rho$ meson states, while diagram (b) reflects the background contribution.}\label{Fey}
\end{figure}

As mentioned in the introduction, due to the conservation of isospin and $G$ parity, both of the $e^+ e^- \to \omega\pi^0$ and $e^+ e^- \to\rho\eta^{\prime}$ reactions are the clean processes for studying $\rho$ meson states.
For the discussed $e^+ e^- \to \omega\pi^0$ and $e^+ e^- \to\rho\eta^{\prime}$ reactions, there exist two mechanisms as shown in Fig. \ref{Fey}. Here,  not only does the virtual photon from the $e^+e^-$ annihilation directly couple with final meson states, but also it can first interact with the intermediate
$\rho$ mesons, which decay into the final meson states.

For calculating the cross section of these reactions, we adopt
 the effective Lagrangian approach.
The involved effective Lagrangians include \cite{Bauer:1975bv,Bauer:1975bw,Kaymakcalan:1983qq,Lin:1999ad,Oh:2000qr,Chen:2011cj}
\begin{eqnarray}
\mathcal{L}_{\rho^*\mathcal{V}\mathcal{P}}&=&g_{\rho^*\mathcal{V}\mathcal{P}}\epsilon_{\mu\nu\alpha\beta}\partial^{\mu}\rho^{*\nu}\partial^{\alpha}\mathcal{V}^{\beta}\mathcal{P},\\
\mathcal{L}_{\mathcal{V}\mathcal{P}\mathcal{P}}&=&g_{\mathcal{V}\mathcal{P}\mathcal{P}}\mathcal{V}^{\mu}\mathcal{P}\overleftrightarrow{\partial}_{\mu}\mathcal{P},\\
\mathcal{L}_{\gamma \mathcal{V}\mathcal{P}}&=&e\epsilon_{\mu\nu\alpha\beta}\partial^{\mu}A^{\nu}\partial^{\alpha}\mathcal{V}^{\beta}\mathcal{P},\\
\mathcal{L}_{\gamma \rho^*}&=&-e\frac{m_{\rho^*}^2}{f_{\rho^*}}\rho^{*\mu}A_{\mu},
\end{eqnarray}
where the $\rho^*$, $\mathcal{V}$, and $\mathcal{P}$ denote the fields of excited $\rho$ meson states, the vector field, and pseudoscalar field, respectively.
With the above effective Lagrangians, relevant amplitudes of $e^+e^-\to \omega\pi^0$ corresponding to diagrams in Fig. \ref{Fey} are written as
\begin{eqnarray}\nonumber\label{amplitude}
\mathcal{M}_{\rho^*_i}&=&\bar{v}(p_2)(ie\gamma^{\mu})u(p_1)\frac{-g_{\mu\nu}}{q^2}(-e\frac{m_{\rho^*_i}^2}{f_{\rho^*_i}})\frac{-g_{\nu\delta}+q_{\nu}q_{\delta}/q^{2}}{q^2-m_{\rho^*_i}^2+im_{\rho^*_i}\Gamma_{\rho^*_i}}\\
&&\times\left[g_{\rho^*_i \omega\pi^0}\epsilon^{\tau\delta\alpha\beta}(-iq_{\tau})(ip_{3\alpha})\right]\varepsilon_{\beta}(p_3),\\
\mathcal{M}_{\rm{Dir}}&=&\bar{v}(p_2)(ie\gamma^{\mu})u(p_1)\frac{-g_{\mu\nu}}{q^2}\left[e\epsilon_{\tau\nu\alpha\beta}(-iq^{\tau})(ip_{3}^{\alpha})\right]\\ \nonumber
&&\times\varepsilon_{\beta}(p_3)\mathcal{F}(s),
\end{eqnarray}
where $p_1$, $p_2$, $p_3$, and $p_4$
are the four-momentum of $e^+$, $e^-$, $\omega$, and $\pi^0$, respectively, and $q=p_1+p_2=p_3+p_4$. The $\mathcal{F}(s)=a\exp\left({-b(\sqrt{s}-\sum_{f} m_f)}\right)/s$ denotes the form factor, {where the $a$ and $b$ are free parameters, which can be determined by fitting experimental data, and the $\sum_{f} m_f$ is the sum for the masses of the final particles.}
The total amplitude of $e^+e^-\to \omega\pi^{0}$ is
\begin{eqnarray}
\mathcal{M}_{\rm{Total}}=\mathcal{M}_{\rm{Dir}}+\sum_{\rho^*_i}\mathcal{M}_{\rho^*_i}e^{i\phi^{\rho^*_i}},
\end{eqnarray}
where the $\phi^{\rho^*_i}$ is the phase angle between the amplitudes from the direct annihilation and intermediate $\rho$ meson state contributions.
With the above amplitudes, the Born cross section of $e^+ e^-\to \omega \pi^0$ can be calculated by
\begin{eqnarray}\label{Icrosssection}
d\sigma=\frac{1}{32\pi s}\frac{|\vec{p}_{3cm}|}{|\vec{p}_{\rm{1cm}}|}\overline{|\mathcal{M}_{\rm{Total}}|^{2}}d\rm{cos}\theta.
\end{eqnarray}
Here, $\theta$ is the scattering angle of an outgoing $\omega$ relative to the direction of the electron beam in the center-of-mass frame, while $\vec{p}_{1cm}$ and $\vec{p}_{\rm{3cm}}$ are the three-momentum of the electron and $\omega$ in the center-of-mass frame, respectively.
The overline in $\overline{|\mathcal{M}_{\rm{Total}}|^{2}}$ indicates the average over the polarizations of $e^+e^-$ in the initial states and the sum over the polarizations of $\omega \pi^0$ in the final states.

Since experiments usually measure $\Gamma_{e^+e^-}\mathcal{B}(\rho^*\to\omega\pi^0)$ rather than the coupling constant $g_{\rho^*\omega\pi^0}/f_{\rho^*}$, we replace $g_{\rho^*\omega\pi^0}/f_{\rho^*}$ in Eq. (\ref{amplitude}) with the function of $\Gamma_{e^+e^-}\mathcal{B}(\rho^*\to\omega\pi^0)$ for the sake of convenience.
With the effective Lagrangians given above, the dilepton and $\omega \pi^0$ decay widths of the discussed $\rho$ mesons can be written as
\begin{eqnarray}
\Gamma_{e^+ e^-}=\frac{e^4 m_{\rho^*_{i}}}{12\pi f_{\rho^{*}_{i}}^{2}} \quad {\rm and}\quad
\Gamma_{\omega\pi^0}=\frac{ g_{\rho^{*}_{i}\omega\pi^0}^2 |\vec{p}_{3\rm{cm}}|^3}{12\pi},
\end{eqnarray}
where the $\vec{p}_{3\rm{cm}}$ is the three-momentum
of $\omega$ in the center-of-mass frame, which can be  expressed as $|\vec{p}_{3\rm{cm}}| =\sqrt{[m_{\rho_i^*}^2 - (m_{\omega} - m_{\pi^0})^2][m_{\rho_i^*}^2 - (m_{\omega} + m_{\pi^0})^2]}/(2 m_{\rho_i^*})$.
So the absolute value of the coupling constant $g_{\rho^*\omega\pi^{0}}/f_{\rho^*}$ can be expressed as
\begin{eqnarray}\label{couplingconstant}
\left|\frac{g_{\rho_{i}^*\omega\pi^0}}{f_{\rho_{i}^*}}\right|=\sqrt{\frac{144\pi^{2}\Gamma_{\rho_{i}^{*}}\Gamma_{e^+e^-}\mathcal{B}(\rho_{i}^{*}\to\omega\pi^0)}{m_{\rho_{i}^{*}}|\vec{p}_{3\rm{cm}}|^3}}.
\end{eqnarray}

{For the reaction $e^+e^-\to \rho\eta^{\prime}\to\eta^{\prime}\pi^-\pi^-$, we can calculate its cross section by replacing the masses of the involved $\omega$ and  $\pi^0$ mesons in Eqs. (\ref{Icrosssection})-(\ref{couplingconstant}) with the corresponding masses of $\rho$ and $\eta^{\prime}$ mesons, and then replacing the $\varepsilon_{\beta}(p_3)$ to $$\frac{-g_{\beta\kappa}+p_{3\beta}p3_{3\kappa}/m_{\rho}^2}{p_{3}^{2}-m_{\rho}^2+im_{\rho}\Gamma_{\rho}}g_{\rho\pi\pi}(ip_{5}^{\kappa}-ip_{6}^{\kappa}),$$ where the $p_5$ and $p_6$ are the four momenta of $\pi^+$ and $\pi^-$, respectively.
By utilizing the partial decay widths of $\rho\to\pi\pi$ listed in PDG \cite{ParticleDataGroup:2020ssz}, we obtain the $g_{\rho\pi\pi}$=6.0 $\rm{GeV}^2$.}

\section{numerical results}\label{section3}

In this section, we study the Born cross sections of $e^+ e^- \to \omega \pi^0$ \cite{BESIII:2020xmw} and $e^+e^-\to\eta^{\prime}\pi^+\pi^-$ \cite{BESIII:2020kpr}.
As a first step, we need to identify the contributions of different intermediate $\rho$ meson states to these two reactions with theoretical support on the spectrum and decay behaviors of these $\rho$ meson states around 2 GeV.
In Table \ref{putin}, we present the information of masses and decay behaviors of these three $\rho$ mesons around 2 GeV.
Here, the strong decay behaviors were estimated by the quark pair creation model \cite{He:2013ttg}, in which the $\rho(1900)$, $\rho(2000)$, and $\rho(2150)$ are assigned as $\rho(3^3S_1)$, $\rho(2^3D_1)$, and $\rho(4^3S_1)$, respectively.
The dilepton widths of vector mesons can be estimated according to the zero-point behavior of their radiative wave functions, which are given in Refs. \cite{He:2013ttg,Wang:2020kte,Godfrey:1985xj}.
The mass of $\rho(1900)$ shown in Table \ref{putin} is adopted from the average value of experimental measurements collected in PDG \cite{ParticleDataGroup:2020ssz} because the measurements of different experiments are almost similar within the range of errors.
However, the $\rho(2000)$ lacks direct experimental measurements, and the different experimental measurements of the mass of $\rho(2150)$ are in a large range from 1910 to 2310 MeV.
In these cases, it becomes meaningless to take the average of the measurements collected in PDG.
Therefore, the masses of $\rho(2000)$ and $\rho(2150)$ are adopted from the values obtained from an analysis of the Regge trajectory \cite{He:2013ttg} in this work.
For the $e^+e^-\to\omega\pi^0$, the contribution of $\rho(2150)$ must be taken into account.
But it is difficult for us to identify the contributions of $\rho(1900)$ and $\rho(2000)$ only from the values of $\Gamma_{e^+e^-}\mathcal{B}(\rho^*\to\omega\pi^0)$ because the value of $\Gamma_{e^+e^-}\mathcal{B}(\rho(2000)\to\omega\pi^0)$ is comparable with lower limit
of $\Gamma_{e^+e^-}\mathcal{B}(\rho(1900)\to\omega\pi^0)$.
Here, we adopt four different schemes to analyze the contributions of $\rho(1900)$ and $\rho(2000)$ in $e^+e^-\to\omega\pi^0$.

\begin{table*}
  \centering
  \caption{The information of $\rho$ meson states involved in the $e^+e^-\to\omega\pi^0$ and $e^+e^-\to\rho\eta^{\prime}$ processes around 2.0 GeV. The $R$ is the parameter in a simple harmonic oscillator wave function. $\Gamma_{e^+e^-}$ is the dilepton decay width that was estimated by Ref. \cite{Wang:2020kte}. $\mathcal{B}(\omega\pi^0)$ and $\mathcal{B}(\rho\eta^{\prime})$ are the branching ratios of $\omega\pi^0$ and $\rho\eta^{\prime}$ modes estimated via the quark pair creation model \cite{He:2013ttg}, respectively . In the last two lines, we give the values of a product of $\Gamma_{e^+e^-}$ and $\mathcal{B}(\omega\pi^0)$, and $\Gamma_{e^+e^-}$ and $\mathcal{B}(\rho\eta^{\prime})$, respectively.}\label{putin}
  \begin{tabular}{ccccccc}
  \toprule[1pt]
  \midrule[1pt]
  State & $\quad$ & $\rho(1900)$ & $\quad$ & $\rho(2150)$ & $\quad$ & $\rho(2000)$\\
  \midrule[1pt]
  Mass (MeV) & $\quad$ & $1890\pm20$ \cite{ParticleDataGroup:2020ssz}& $\quad$ & $2160$ \cite{He:2013ttg} & $\quad$ & $2040$ \cite{He:2013ttg} \\
  $R$ (GeV$^{-1}$) \cite{He:2013ttg,Wang:2020kte} & $\quad$ & $3.8\sim 4.3$ & $\quad$ & $4.5\sim 5.0$ & $\quad$ & $4.3\sim4.8$ \\
  $\Gamma_{e^+e^-}$ (eV) \cite{Wang:2020kte}& $\quad$ & $166.31\sim213.91$ & $\quad$ & $78.72\sim 96.80$ & $\quad$ & $28.74\sim 16.74$\\
  $\mathcal{B}(\omega \pi^0)$ ($10^{-3}$) \cite{He:2013ttg} & $\quad$ & $0.18\sim 24.79 $ & $\quad$  & $18.22\sim 88.76$ & $\quad$ & $15.01\sim19.65$\\
  $\mathcal{B}(\rho \eta^{\prime})$ ($10^{-3}$) \cite{He:2013ttg} & $\quad$ & $6.85\times10^{-4}\sim 2.73$ & $\quad$  & $0.94\sim2.65$ & $\quad$ & $5.55\sim15.78$\\
  $\Gamma_{e^+e^-}\mathcal{B}(\omega\pi^0)$ (eV) & $\quad$ &$0.03\sim 5.31$ & $\quad$ & $1.76\sim 6.99$ & $\quad$ & $0.23\sim 0.38$\\
  $\Gamma_{e^+e^-}\mathcal{B}(\rho\eta^{\prime})$ (eV) & $\quad$ &$0.01\times10^{-2}\sim0.45$ & $\quad$ & $0.10\sim 0.20$ & $\quad$ & $0.11\sim 0.24$\\
\midrule[1pt]
\bottomrule[1pt]
\end{tabular}
\end{table*}

\begin{figure*}[htb]
  \centering
  \begin{tabular}{ccc}
  \includegraphics[width=220pt]{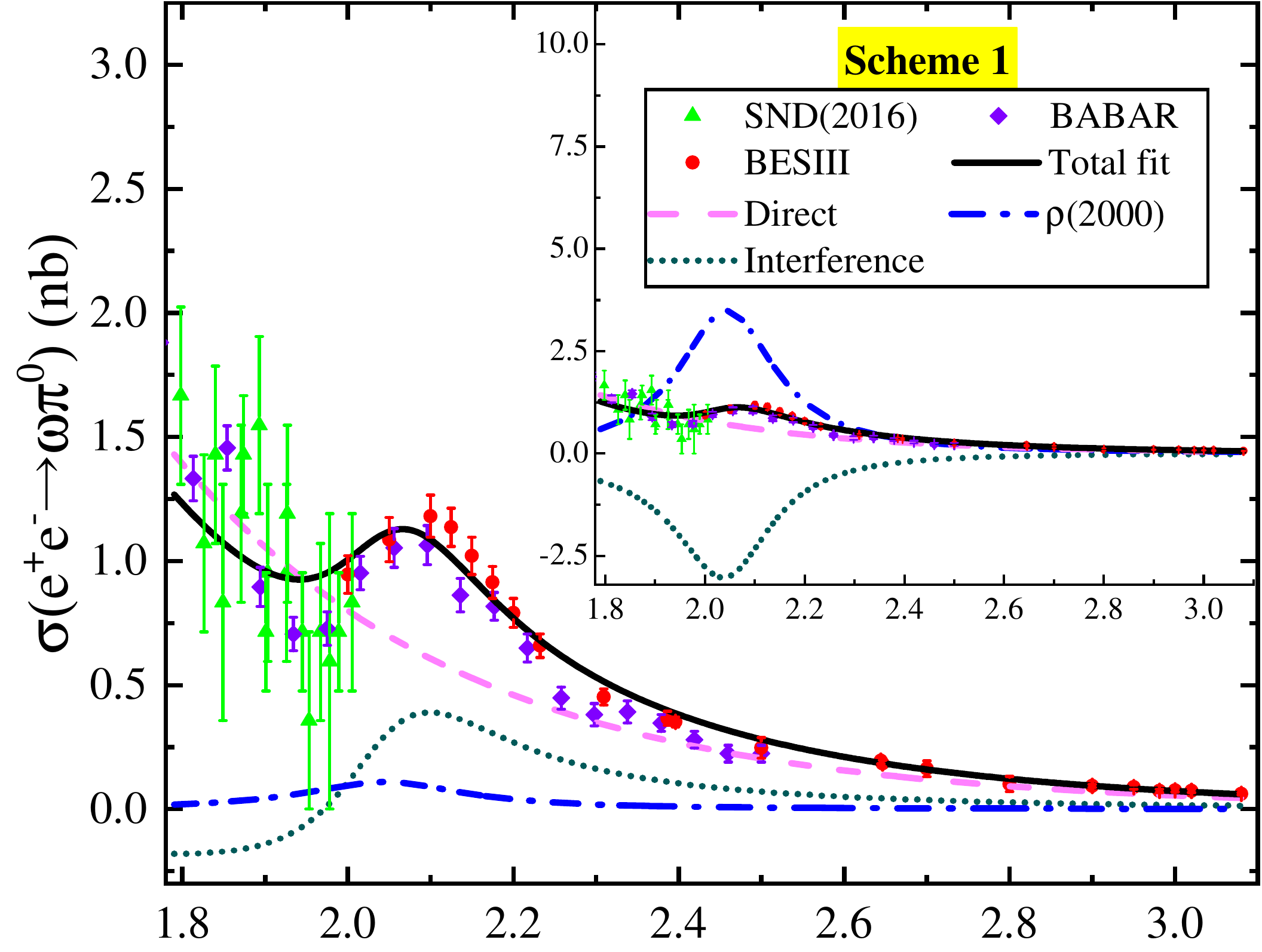} & $\quad$ & \includegraphics[width=220pt]{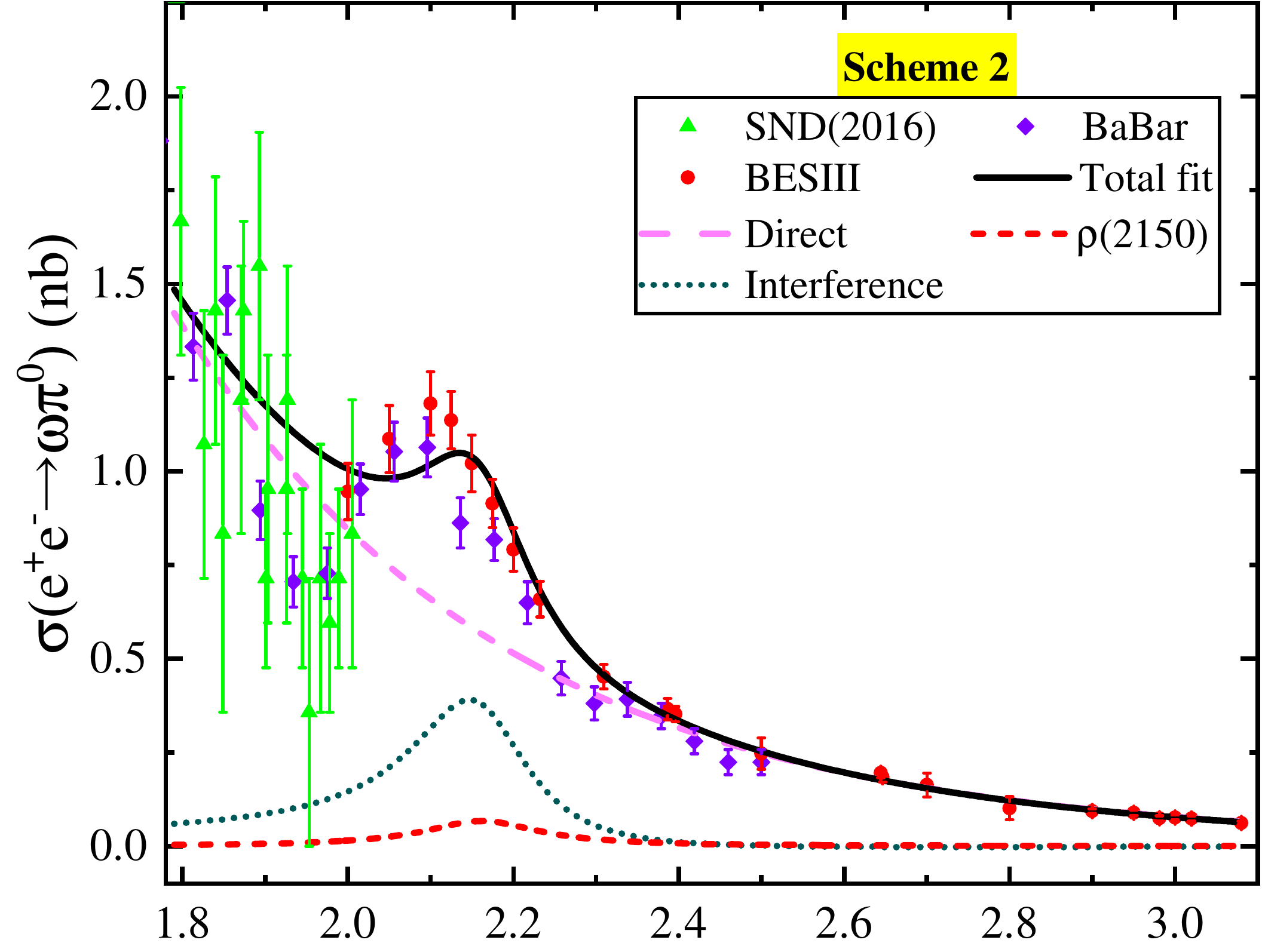} \\
  \includegraphics[width=220pt]{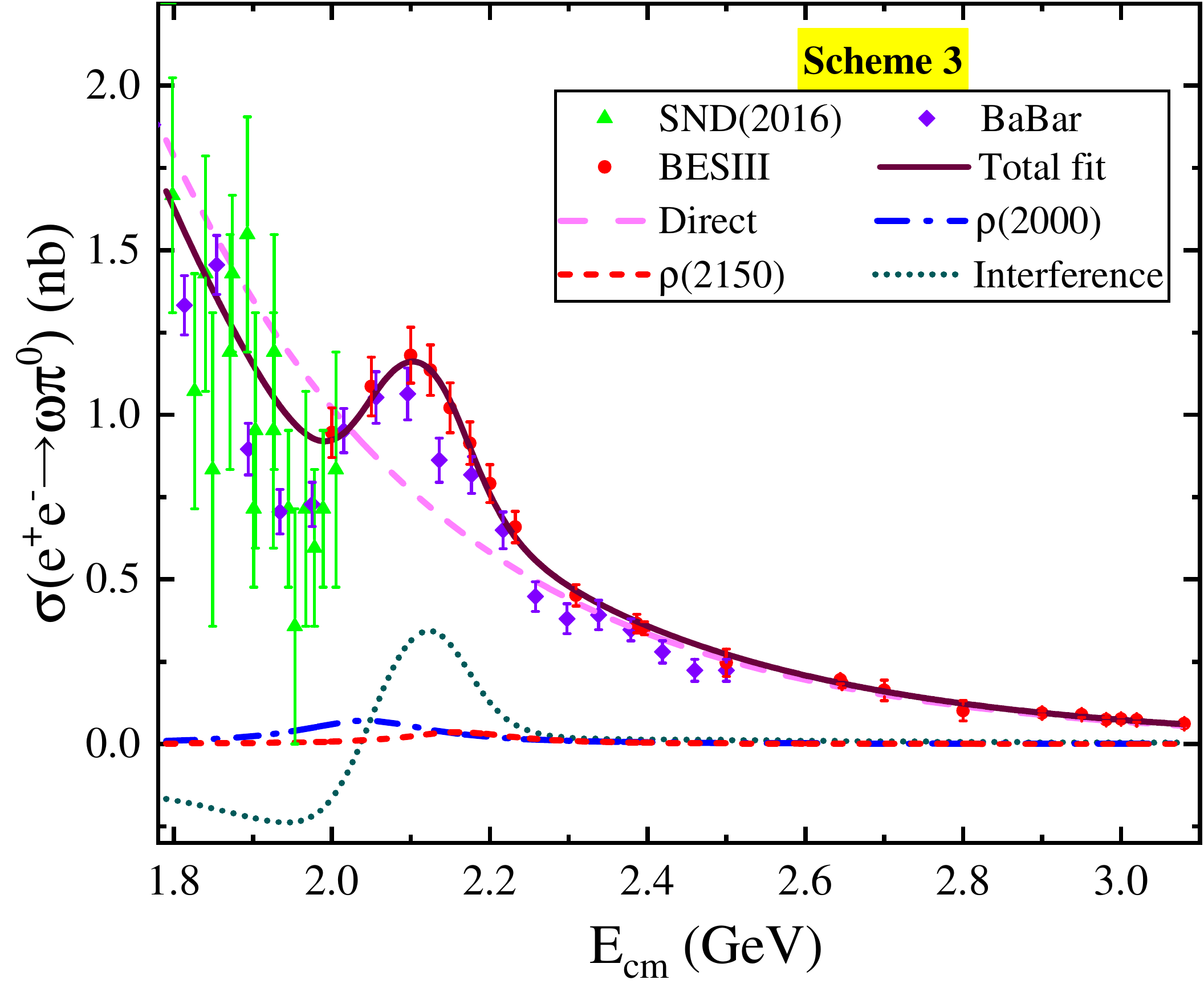} & $\quad$ & \includegraphics[width=220pt]{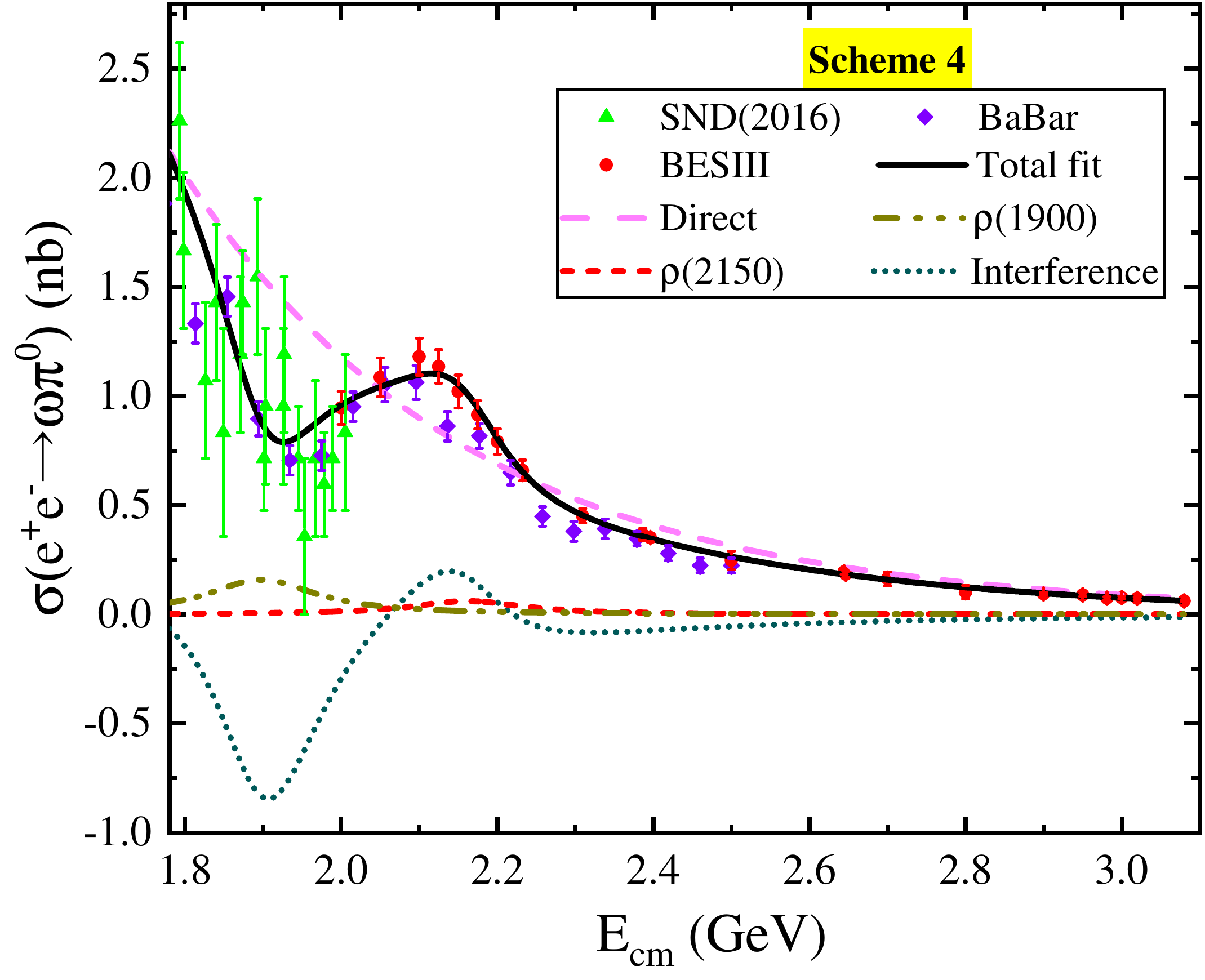} \\
  \end{tabular}
  \caption{The fitted results of the Born cross sections of $e^+e^-\to\omega\pi^0$ measured by the BESIII \cite{BESIII:2020xmw} (red dots with error bars) and SND \cite{Achasov:2016zvn} (green dots with error bars) collaborations with different schemes. For comparison, we also present the experimental data  measured by the BaBar \cite{BaBar:2017zmc} (violet dots with error bars) Collaboration. In the scheme 1 of $e^+e^-\to\omega\pi^0$, the main figure is a result of constructive interference of $\rho(2000)$, and the insert is a result of destructive interference.}\label{figurefittingresultop}
\end{figure*}

In our fitting process, we take the input masses of the $\rho(1900)$, $\rho(2000)$, and $\rho(2150)$ from Table \ref{putin}, and we regard the widths, the relevant combined branching ratios $\Gamma_{e^+e^-}\mathcal{B}(\rho^*_{i}\to{\omega\pi^0})$, the phases $\phi^{\rho^*_i}_1$, and the parameters in the form factor $a_1$, $b_1$ as free parameters, which are determined by fitting experimental data of the Born cross sections of $e^+e^-\to\omega\pi^0$ measured by the BESIII \cite{BESIII:2020xmw} and SND \cite{Achasov:2016zvn} collaborations.
{It needs to be explained here that the experimental data given by the SND Collaboration \cite{Achasov:2016zvn} are the cross sections of $e^+e^-\to\omega\pi^0\to\gamma\pi^0\pi^0$, and then the cross sections of $e^+e^-\to\omega\pi^0$ can be obtained by dividing by the branching ratio of the $\omega$ decay to $\gamma\pi^0$.
Here, the branching ratio of $e^+e^-\to\omega\pi^0$ is adopted by the value $8.40\%$, which is given in PDG \cite{ParticleDataGroup:2020ssz}.
}
The fitting parameters of different schemes are presented in Table \ref{parametersfittingresltsop}.
Applying the central values of fitted parameters, we can plot the Born cross section of $e^+e^-\to \omega \pi^0$ as a function of center-of-mass energy, and show them in Fig. \ref{figurefittingresultop}.

\begin{table*}[htb]
  \centering
  \caption{The parameters obtained by fitting the Born cross sections of $e^+e^-\to\omega \pi^0$ measured by the BESIII \cite{BESIII:2020xmw} and SND \cite{Achasov:2016zvn} collaboration .}\label{parametersfittingresltsop}
  \begin{tabular}{ccccccccc}
  \toprule[1pt]
  \midrule[1pt]
  Parameters &  $\quad$ & Scheme 1   & $\quad$ & Scheme 2   &$\quad$ & Scheme 3   &$\quad$ & Scheme 4 \\
  \midrule[1pt]
  $\Gamma_{\rho(1900)}$ (MeV) & $\quad$ & ...  & $\quad$ &...  & $\quad$ & ...  & $\quad$   & $169\pm15$\\
  $\Gamma_{\rho(2000)}$ (MeV) & $\quad$ &$238\pm22$ & $\quad$ &...  & $\quad$ &$206\pm29$ & $\quad$ &  ... \\
  $\Gamma_{\rho(2150)}$ (MeV) & $\quad$ &  ... & $\quad$ & $171\pm23$  & $\quad$& $169\pm23$ & $\quad$ & $177\pm19$\\
  $\Gamma_{e^+e^-}\mathcal{B}(\rho(1900)\to\omega\pi^0)$ (eV)  & $\quad$ &  ... & $\quad$ & ... & $\quad$ & ... & $\quad$ & $6.59\pm1.91$\\
  $\Gamma_{e^+e^-}\mathcal{B}(\rho(2000)\to\omega\pi^0)$ (eV) & $\quad$&  \begin{tabular}{c}$7.37\pm1.52$ \footnote{Solution of constructive interference in scheme 1.}\\$233.52\pm4.87$ \footnote{Solution of destructive interference in scheme 1.}\end{tabular}   & $\quad$& ...   & $\quad$& $4.15\pm0.33$ & $\quad$& ... \\
  $\Gamma_{e^+e^-}\mathcal{B}(\rho(2150)\to\omega\pi^0)$ (eV)  & $\quad$ & ... &  & $\quad$ $3.65\pm0.44$  & $\quad$ & $1.92\pm0.42$  & $\quad$ & $3.40\pm0.97$\\
  $a_1$ ($\rm{GeV}^{-2}$)  & $\quad$ & $2.93\pm0.21$  & $\quad$ & $2.58\pm0.02$   & $\quad$ & $3.36\pm0.28$  & $\quad$ & $3.38\pm0.29$ \\
  $b_1$ ($\rm{GeV}^{-1}$) & $\quad$ & $0.68\pm0.01$  & $\quad$ & $0.53\pm0.01$  & $\quad$ & $0.69\pm0.05$  & $\quad$ & $0.63\pm0.05$ \\
  $\phi^{\rho(1900)}_{1}$ (rad) & $\quad$&  ...  & $\quad$& ... & $\quad$&  ...  & $\quad$  & $4.31\pm0.11$ \\
  $\phi^{\rho(2000)}_{1}$ (rad) & $\quad$ & \begin{tabular}{c}$0.52\pm0.09$ $^a$ \\$4.63\pm0.03$ $^b$ \end{tabular} & $\quad$ & ... & $\quad$ &  $5.96\pm0.17$  & $\quad$ & ... \\
  $\phi^{\rho(2150)}_{1}$ (rad) & $\quad$ & ...  & $\quad$ & $1.78\pm0.08$ & $\quad$ & $2.62\pm0.11$ & $\quad$ &  $2.10\pm0.07$\\
  $\chi^2/\rm{d.o.f.}$  & $\quad$ & 1.25  & $\quad$& 1.13 & $\quad$ & 0.72  & $\quad$ &  0.67 \\
\midrule[1pt]
\bottomrule[1pt]
\end{tabular}
\end{table*}

From the measured resonance parameters of the enhancement structure near 2034 MeV observed in $e^+e^-\to\omega\pi^0$, it seems natural to interpret this enhancement structure as a contribution from $\rho(2000)$.
In Refs. \cite{Li:2021qgz,Yu:2021ggd}, the authors interpreted this structure as $\rho(2000)$ based on a study of the mass spectrum and strong decay width of $\rho$ meson states.
Thus, we try to reproduce the experimental data of the Born cross section of the reaction $e^+e^-\to\omega\pi^0$ only by using $\rho(2000)$ as the intermediate resonance state in the scheme 1.
As seen in Fig. \ref{figurefittingresultop}, we can find that the experimental data measured by the BESIII Collaboration can barely be reproduced in the scheme 1.
However, the fitting solution of constructive interference of $\Gamma_{e^+e^-}\mathcal{B}(\rho(2000)\to\omega\pi^0)$ is {$7.37\pm1.52$ eV, which is about 20 times as large as theoretical estimate $0.23\sim0.38$ eV presented in Table \ref{putin}}, and another solution of destructive interference is {$233.52\pm4.87$}, which is about three orders of magnitude larger than theoretical estimate.
Therefore, our conclusion from the scheme 1 is that the enhancement structure observed near 2034 MeV in $e^+ e^-\to \omega\pi^0$ cannot be interpreted as a contribution only from the $\rho(2000)$.

In the scheme 2, we only consider the $\rho(2150)$ as an intermediate resonance in the $e^+e^-\to \omega \pi^0$.
Here, we need to point out that all the schemes virtually have multiple solutions, but the solutions of destructive interference deviate greatly from the theoretical estimates, so we do not mention multiple solutions in the following discussion.
As shown in Fig. \ref{figurefittingresultop}, this scheme not only cannot well describe the experimental data above 2 GeV measured by the BESIII Collaboration, but also fails to describe the experimental data measured by SND \cite{Achasov:2016zvn} and BaBar \cite{BaBar:2017zmc}.
Combining the results of the schemes 1 and 2, we can conclude that we cannot reproduce the experimental data of the Born reaction cross section of $e^+e^- \to \omega\pi^0$ well with a single resonance fitting.

In the scheme 3, we further assume that the enhancement structure observed in $e^+e^-\to\omega\pi^0$ process is the contribution from $\rho(2000)$ and $\rho(2150)$.
Although the experimental data above 2 GeV measured by the BESIII Collaboration can be reproduced by the scheme 3, the fitting value of $\Gamma_{e^+e^-}\mathcal{B}(\rho(2000)\to\omega\pi^0)$ is {$4.15\pm0.33$} eV, which is still about one order of magnitude larger than the theoretical estimate.
{On the other hand, the line shape in the energy range from 1.8 GeV to 2 GeV is not consistent with the experimental data measured by SND \cite{Achasov:2016zvn} and BaBar \cite{BaBar:2017zmc}}.

In the scheme 4, we consider the contributions from $\rho(1900)$ and $\rho(2150)$ in $e^+e^-\to\omega\pi^0$ process to reproduce experimental data.
From the fitting result of the scheme 4 in Fig. \ref{figurefittingresultop}, we can see that this scheme not only can well reproduce the experimental data above 2 GeV measured by the BESIII Collaboration, but also can well describe the experimental data measured by SND \cite{Achasov:2016zvn} and BaBar \cite{BaBar:2017zmc} in the range of 1.8 to 2 GeV.
Comparing all the schemes, we have reason to guess that the dip near 1.9 GeV shown in SND \cite{Achasov:2016zvn} and BaBar \cite{BaBar:2017zmc} possibly corresponds to the contribution of $\rho(1900)$.
This conclusion can be tested by more precise measurement in the future with more accurate and richer experimental data.
As seen in the fitted parameters of the scheme 4 shown in Table \ref{parametersfittingresltsop}, the values of $\Gamma_{e^+e^-}\mathcal{B}(\rho(1900)\to\omega\pi^0)$ and $\Gamma_{e^+e^-}\mathcal{B}(\rho(2150)\to\omega\pi^0)$ are {$6.59\pm1.91$ eV and $3.40\pm0.97$} eV, respectively, both of which are within the error range of theoretical estimates shown in Table \ref{putin}.
Obviously, there is no clear evidence of $\rho(2000)$ in $e^+e^-\to\omega\pi^0$ based on our analysis, and the theoretical estimate of $\Gamma_{e^+e^-}\mathcal{B}(\rho^*\to\omega\pi^0)$ also implies that $\rho(2000)$ does not make a dominant contribution in $e^+e^-\to\omega\pi^0$.
Finally, we find that the $\chi^2/\rm{d.o.f.}$ of scheme 4 is the smallest of all the schemes.
Based on the analysis of the above four schemes, we can conclude that the enhancement structure near 2034 MeV observed in $e^+e^-\to \omega\pi^0$ by the BESIII Collaboration cannot be interpreted as a single resonance contribution from the $\rho(2000)$, but it is dominated by the contributions of the $\rho(1900)$ and $\rho(2150)$.

{In order to further test the role of $\rho(2000)$ in $e^+e^-\to\omega\pi^0$, we add the contribution of the $\rho(2000)$ in a $3R$ fitting scheme to fit the experimental data. The fitting results are shown in Fig. 4, and the corresponding parameters are summarized in Table III. It can be seen that the $\chi^2/\rm{d.o.f.}=0.66$ in a $3R$ fitting scheme can be obtained.
Compared with the value of $\chi^2/\rm{d.o.f.}=0.67$ in scheme 4, the fitting quality is not significantly improved after adding the contribution of $\rho(2000)$.
On the other hand, the fitting value of $\Gamma_{e^+e^-}\mathcal{B}(\rho(2000)\to\omega\pi^0)$ is $0.48\pm0.28$ eV, which is still about one order of magnitude smaller than the values of $\Gamma_{e^+e^-}\mathcal{B}(\rho(1900)\to\omega\pi^0)$ and $\Gamma_{e^+e^-}\mathcal{B}(\rho(2150)\to\omega\pi^0)$.
It is worth noting that this fitted $\Gamma_{e^+e^-}\mathcal{B}(\rho(2000)\to\omega\pi^0)$=$0.48\pm0.28$ eV is also consistent with our theoretical estimates of $0.23\sim0.38$ eV.
Therefore, it proves that the observed enhancement structure near 2.0 GeV in $e^+e^-\to\omega\pi^0$ is dominantly produced by the contributions of the $\rho(1900)$ and $\rho(2000)$ again, while the contribution of the $\rho(2000)$ is insignificant.}

\begin{table}[htb]
  \centering
  \caption{The parameters obtained by fitting the Born cross sections of $e^+e^-\to\omega \pi^0$ measured by the BESIII \cite{BESIII:2020xmw} and SND \cite{Achasov:2016zvn} Collaborations with the 3R fitting scheme.}\label{3RfitT}
  \begin{tabular}{ccc}
  \toprule[1pt]
  \midrule[1pt]
  Parameters &  $\quad$ & Values \\
  \midrule[1pt]
  $\Gamma_{\rho(1900)}$ (MeV) & $\quad$ &$175\pm14$\\
  $\Gamma_{\rho(2000)}$ (MeV) & $\quad$ & $202\pm16$ \\
  $\Gamma_{\rho(2150)}$ (MeV) & $\quad$ & $183\pm16$\\
  $\Gamma_{e^+e^-}\mathcal{B}(\rho(1900)\to\omega\pi^0)$ (eV)  & $\quad$& $0.48\pm0.28$\\
  $\Gamma_{e^+e^-}\mathcal{B}(\rho(2150)\to\omega\pi^0)$ (eV)  & $\quad$ &  $2.82\pm0.61$\\
  $a_1$ ($\rm{GeV}^{-2}$) &  $\quad$ & $3.40\pm0.30$\\
  $b_1$ ($\rm{GeV}^{-1}$) &  $\quad$ & $0.64\pm0.02$\\
  $\phi^{\rho(1900)}_{1}$ (rad) & $\quad$ & $4.19\pm0.20$\\
  $\phi^{\rho(2000)}_{1}$ (rad) & $\quad$ & $0.23\pm0.12$\\
  $\chi^2/\rm{d.o.f.}$ & $\quad$ & 0.66\\
\midrule[1pt]
\bottomrule[1pt]
\end{tabular}
\end{table}

\begin{figure}[htb]
  \centering
  \begin{tabular}{c}
  \includegraphics[width=220pt]{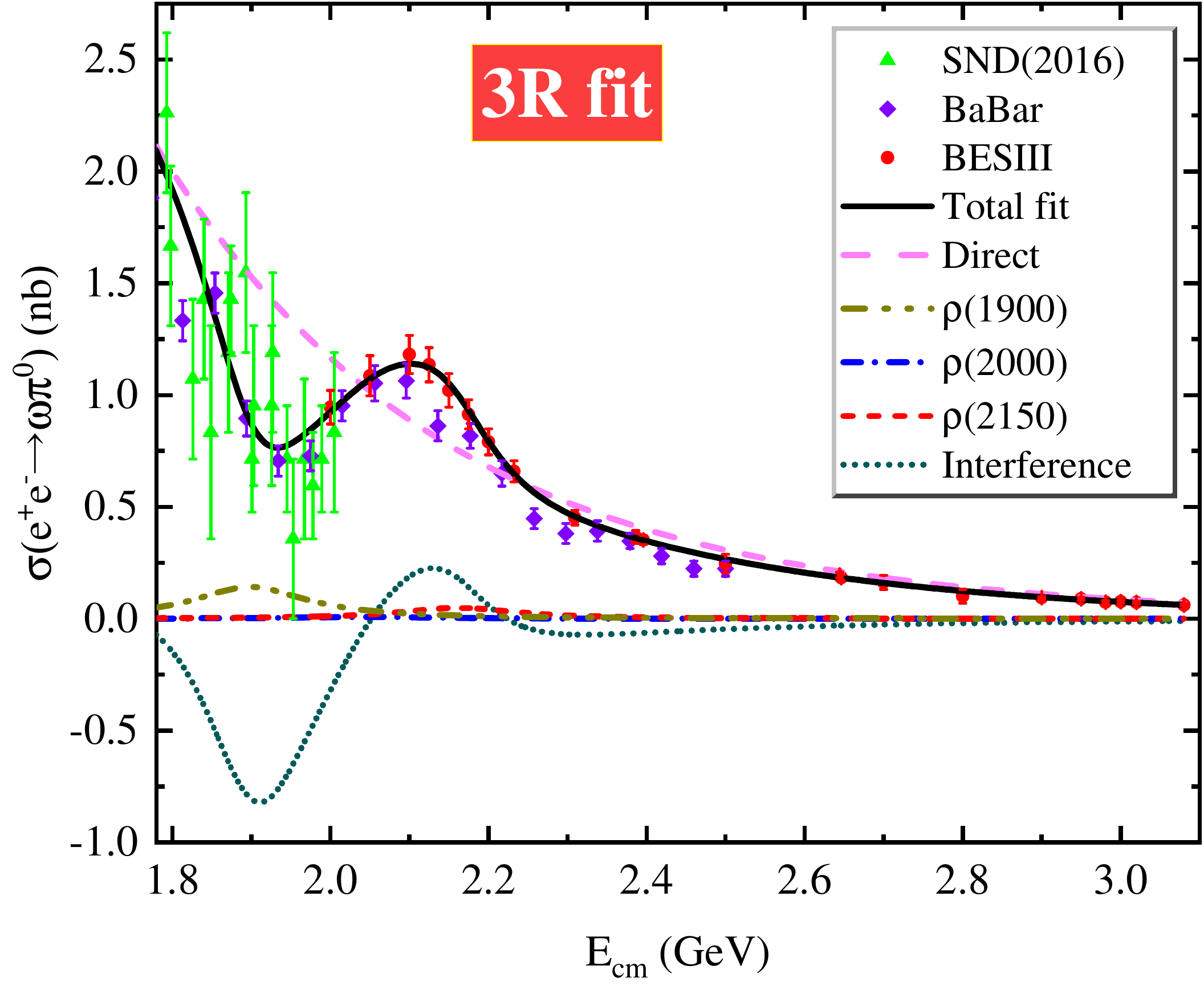} \\
  \end{tabular}
  \caption{The fitted results of the Born cross sections of $e^+e^-\to\omega\pi^0$ measured by the BESIII \cite{BESIII:2020xmw} (red dots with error bars) and SND \cite{Achasov:2016zvn} (green dots with error bars) Collaborations with the 3R fitting scheme.}\label{3RfitF}
\end{figure}

\begin{figure*}[htb]
  \centering
  \begin{tabular}{ccc}
  \includegraphics[width=220pt]{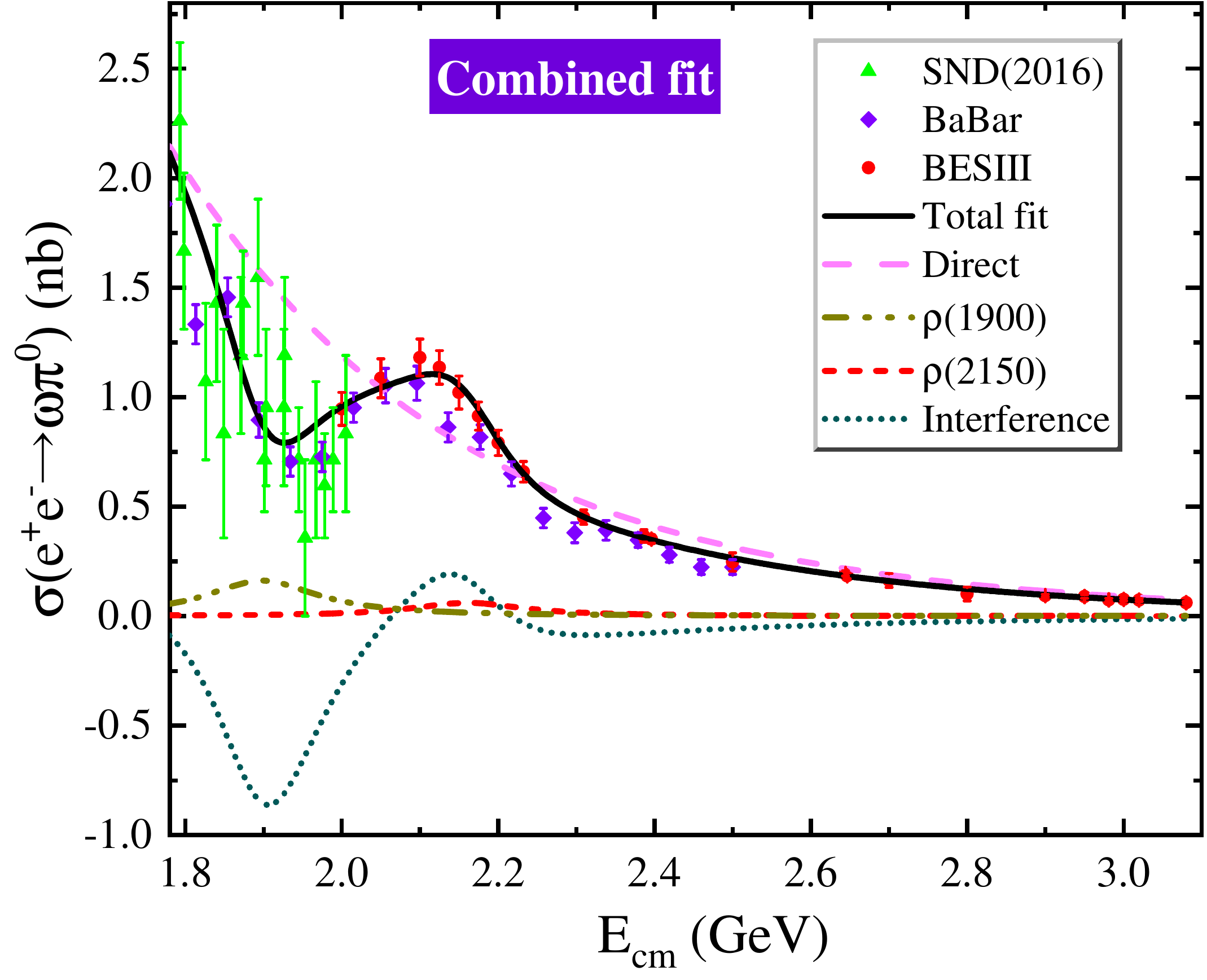} & $\quad$ & \includegraphics[width=220pt]{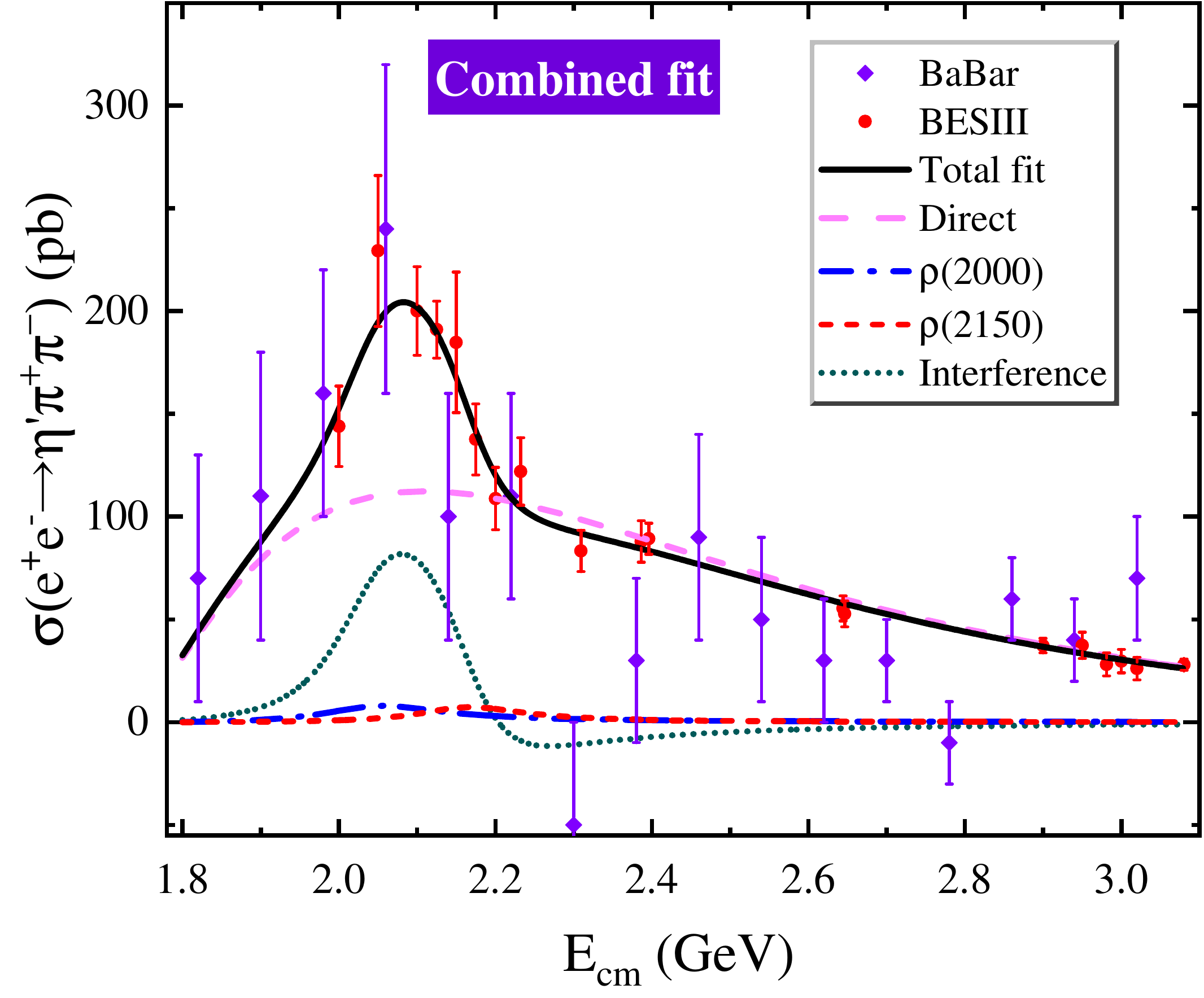} \\
  \end{tabular}
  \caption{The combined fit to the precise Born cross sections of $e^+e^-\to\omega\pi^0$ \cite{BESIII:2020xmw,Achasov:2016zvn} and $e^+e^-\to\eta^{\prime}\pi^+\pi^-$ \cite{BESIII:2020kpr}. For comparison, we also present the experimental data  measured by the BaBar \cite{BaBar:2017zmc,BaBar:2007qju} Collaboration.}
 \label{figurefittingresultcf}
\end{figure*}

Following the analysis for $e^+e^-\to\omega\pi^0$, we search for the possibility of multi-$\rho$-state dominance in the reaction $e^+e^-\to\rho\eta^{\prime}$ too by considering the value of $\Gamma_{e^+e^-}\mathcal{B}(\rho^*\to\rho\eta^{\prime})$. As seen in Table \ref{putin}, the contribution of both $\rho(2000)$ and $\rho(2150)$ are significant and
although the contribution from $\rho(1900)$ may be comparable to $\rho(2000)$ and $\rho(2150)$ in $e^+e^-\to\rho\eta^{\prime}$, its contribution is obviously suppressed in the energy region of the enhancement structure because its mass is far from the enhancement structure position of 2111 MeV. Hence, we can conclude that $\rho(2000)$ and $\rho(2150)$ are dominant for $e^+e^-\to\rho\eta^{\prime}$.
In order to verify this assumption and understand the enhancement structure near 2111 MeV observed in $e^+e^-\to\eta^{\prime}\pi^+\pi^-$ by the BESIII Collaboration, we perform a simultaneous fit to the experimental data of the Born cross sections of $e^+e^-\to\omega\pi^0$ and $e^+e^-\to \eta^{\prime}\pi^+\pi^-$ above 2 GeV measured by the BESIII Collaboration.
The reason why we use the combined fitting scheme is that it can restrict the free resonance parameters more strongly.
From Fig. \ref{figurefittingresultcf}, we can see that the enhancement structure near 2111 MeV observed in $e^+e^-\to\eta^{\prime}\pi^+\pi^-$ by the BESIII Collaboration can be reproduced by the resonance contributions from $\rho(2000)$ and $\rho(2150)$.
From the values of fitted parameters shown in Table \ref{parametersfittingresltscf}, we can see that both of the fitted results of $\Gamma_{e^+e^-}\mathcal{B}(\rho(2000)\to\rho\eta^{\prime})$ and $\Gamma_{e^+e^-}\mathcal{B}(\rho(2150)\to\rho\eta^{\prime})$ are within the theoretical prediction ranges of $0.10\sim0.20$ and $0.11\sim0.24$, respectively.
On the other hand, current experimental data show no signal of enhancement structure in the range of 1.8 to 2.0 GeV, which indicates that it is reasonable to ignore the contribution from $\rho(1900)$ in the process of $e^+e^-\to\rho\eta^{\prime}$.
According to our analysis, we can draw a conclusion that the enhancement structure near 2111 MeV observed in $e^+e^-\to \eta^{\prime}\pi^+\pi^-$ by the BESIII Collaboration can be interpreted as an interference structure from the $\rho(2000)$ and $\rho(2150)$.
As a $D$-wave $\rho$ meson state, the dilepton width of the $\rho(2000)$ is generally suppressed, which leads to the difficulty of searching for the $\rho(2000)$ in $e^+e^-$ collision experiments.
However, our analysis shows that the contributions of the $\rho(2000)$ and $\rho(2150)$ are comparable in the $e^+e^-\to\rho\eta^{\prime}$ reaction.
Therefore, $e^+e^-\to\rho\eta^{\prime}$ may be a golden reaction to study $\rho(2000)$.
{According to the values of $\Gamma_{e^+e^-}\mathcal{B}(\rho(2000)\to\omega\pi^0)$ and $\Gamma_{e^+e^-}\mathcal{B}(\rho(2000)\to\rho\eta^{\prime})$, although the absolute contribution of the $\rho(2000)$ in $e^+e^-\to\omega\pi^0$ and $e^+e^-\to\rho\eta^{\prime}$ are similar to each other, the total cross section near 2 GeV of $e^+e^-\to\omega\pi^0$ is about one order of magnitude larger than $e^+e^-\to\rho\eta^{\prime}$, which means that the relative contribution of $\rho(2000)$ in $e^+e^-\to\rho\eta^{\prime}\to\eta^{\prime}\pi^+\pi^-$ is larger than that in $e^+e^-\to\omega\pi^0$.
Therefore, we conclude that $e^+e^-\to\rho\eta^{\prime}\to\eta^{\prime}\pi^+\pi^-$ is a more appropriate reaction to study $\rho(2000)$.}

\begin{table}[htb]
  \centering
  \caption{The parameters of the combined fit to the experimental data of the Born cross sections of $e^+e^-\to\omega \pi^0$ \cite{BESIII:2020xmw,Achasov:2016zvn} and $e^+e^-\to\eta^{\prime}\pi^+\pi^-$ \cite{BESIII:2020kpr} processes. Here, subscripts 1 and 2 indicate the corresponding parameters belonging to $e^+ e^- \to \omega \pi^0$ and $e^+e^-\to\eta^{\prime}\pi^+\pi^-$, respectively.}\label{parametersfittingresltscf}
  \begin{tabular}{ccc}
  \toprule[1pt]
  \midrule[1pt]
  Parameters & $\quad$ & Values   \\
  \midrule[1pt]
  $\Gamma_{\rho(1900)}$ (MeV) &  $\quad$ &$173\pm18$\\
  $\Gamma_{\rho(2000)}$ (MeV) &  $\quad$ & $194\pm38$ \\
  $\Gamma_{\rho(2150)}$ (MeV) &  $\quad$ &  $175\pm22$\\
  $\Gamma_{e^+e^-}\mathcal{B}(\rho(1900)\to\omega\pi^0)$ (eV)& $\quad$ & $6.88\pm1.92$\\
  $\Gamma_{e^+e^-}\mathcal{B}(\rho(2000)\to\omega\pi^0)$ (eV)& $\quad$ & $3.37\pm0.79$\\
  $a_1$ ($\rm{GeV}^{-2}$) & $\quad$ & $3.41\pm0.33$\\
  $b_1$ ($\rm{GeV}^{-1}$) & $\quad$ & $0.64\pm0.04$\\
  $\phi^{\rho(1900)}_{1}$ (rad)& $\quad$ & $4.34\pm0.14$\\
  $\phi^{\rho(2150)}_{1}$ (rad)& $\quad$ & $2.11\pm0.14$\\
  $\Gamma_{e^+e^-}\mathcal{B}(\rho(2000)\to\rho\eta^{\prime})$ (eV) & $\quad$  & $0.44\pm0.21$\\
  $\Gamma_{e^+e^-}\mathcal{B}(\rho(2150)\to\rho\eta^{\prime})$ (eV) & $\quad$  & $0.51\pm0.27$\\
  $a_2$ ($\rm{GeV}^{-2}$) & $\quad$  & $1.78\pm0.15$\\
  $b_2$ ($\rm{GeV}^{-1}$) & $\quad$  & $0.64\pm0.06$\\
  $\phi^{\rho(2000)}_2$ (rad)& $\quad$  & $6.17\pm0.29$\\
  $\phi^{\rho(2150)}_2$ (rad)& $\quad$  & $4.58\pm0.42$\\
  $\chi^2/\rm{d.o.f.}$ & $\quad$ & 0.74\\
\midrule[1pt]
\bottomrule[1pt]
\end{tabular}
\end{table}

\begin{figure}[htb]
  \centering
  \begin{tabular}{c}
  \includegraphics[width=220pt]{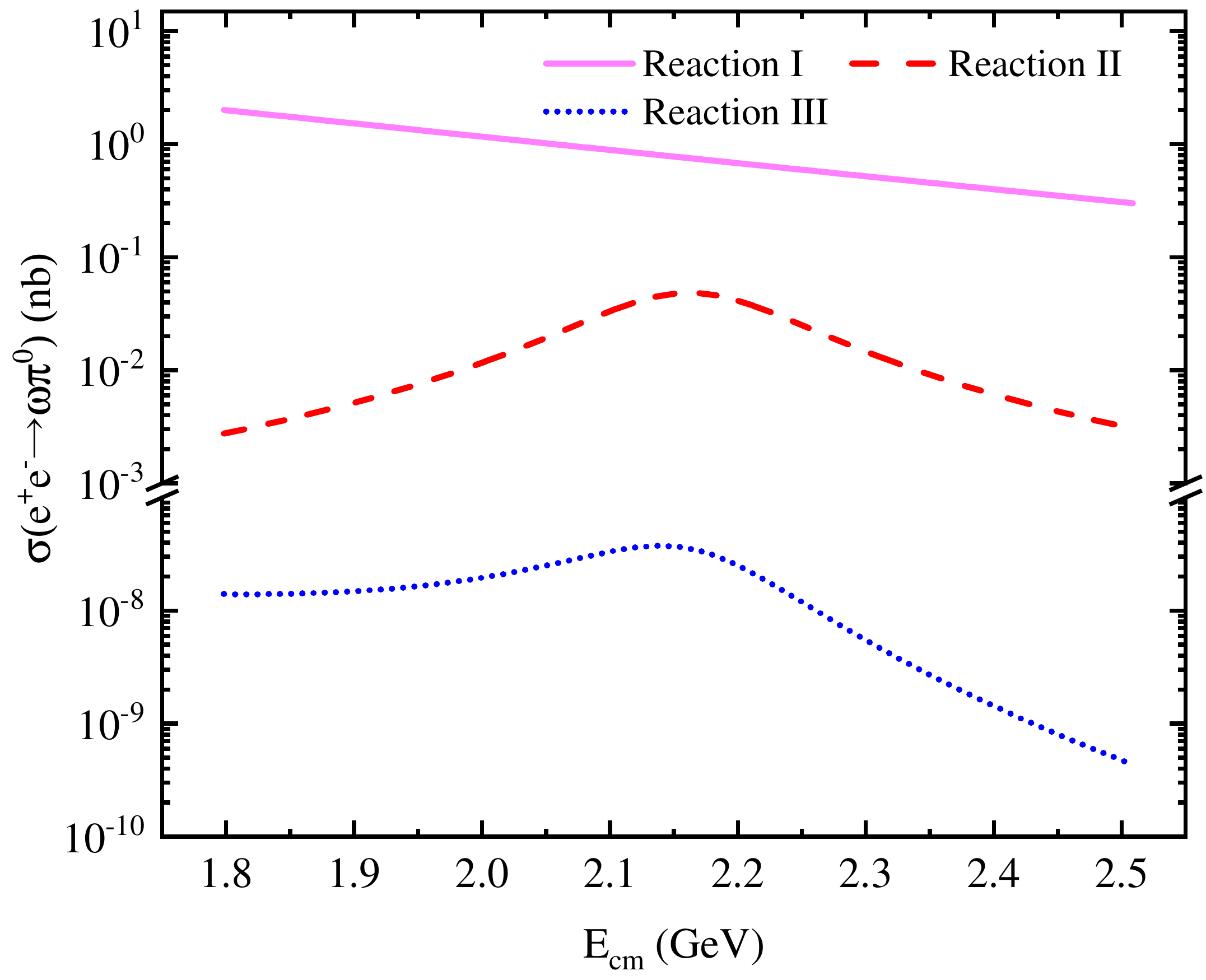}\\
  \end{tabular}
  \caption{Cross section sizes of $e^+e^-\to\omega\pi^0$ are from different reactions, where reaction I, reaction II, and reaction III represent $e^+e^-\to\gamma^*\to\omega\pi^0$, $e^+e^-\to\gamma^*\to\rho(2150)\to\omega\pi^0$ and $e^+e^-\to\gamma^*\to\rho(2150)\to\gamma^*\to\omega\pi^0$, respectively.}\label{Compare}
  \end{figure}

We also present the individual contributions to the Born cross sections in Figs. \ref{figurefittingresultop}, \ref{3RfitF} and \ref{figurefittingresultcf}, we can find that the resonance contribution is obviously smaller than the direct term, but the interference effect is very important to reproduce the experimental data.
{Because the direct process provides a dominant contribution in both of $e^+e^-\to\omega\pi^0$ and $e^+e^-\to\rho\eta^{\prime}$, one expect that it may enhance the contribution from reaction $e^+e^-\to\gamma^*\to\rho^*\to\gamma^* \to \omega\pi^0$ or $\rho \eta^{\prime}$. As an example, we have calculated the cross sections of $e^+e^-\to\gamma^*\to\omega\pi^0$ (reaction I), $e^+e^-\to\gamma^*\to\rho(2150)\to \omega\pi^0$ (reaction II) and $e^+e^-\to\gamma^*\to\rho(2150)\to\gamma^*\to\omega\pi^0$ (reaction III) by using the parameters in Table \ref{parametersfittingresltscf}, whose results are shown in Fig. \ref{Compare}.
It can be seen from Fig. \ref{Compare} that the cross section of $e^+e^-\to\gamma^*\to\rho(2150)\to\gamma^*\to\omega\pi^0$ is much smaller than those of two other mechanisms. The reason for this is that there exist a suppression vertex of electromagnetic interaction.
Therefore, it is reasonable to not consider the contribution of $e^+e^-\to\gamma^*\to\rho(2150)\to\gamma^*\to\omega\pi^0$ or $\rho \eta^{\prime}$ in our calculation.}
The interference effect considerably changes the line shape of the Born cross sections, which makes the original two resonance states to appear as one structure in the experimental data.
This is the important reason why the resonance parameters of $\rho$ meson states around 2.0 GeV are so poorly measured.
In this case, it is not sufficient to understand a structure observed by experiments only by the resonance parameters, and their production and decay behaviors provide more information for us.
On the other hand, the theoretical studies of the spectrum and decay behaviors of $\rho$ meson states around 2 GeV greatly help us identify the contributions of the $\rho$ meson states to these two processes of $e^+e^-\to\omega\pi^0$ and $e^+e^-\to\rho\eta^{\prime}$.
We believe that the resonance parameters of $\rho(1900)$, $\rho(2000)$, and $\rho(2150)$ can be established by more abundant experimental data of $e^+e^-\to\omega\pi^0$ and $e^+e^-\to\rho\eta^{\prime}$ and more theoretical researches in the future.

\section{summary}\label{section4}

Recently, the BESIII Collaboration measured the Born cross sections for the processes $e^+e^-\to \omega \pi^0$ \cite{BESIII:2020xmw} and $e^+e^-\to \rho\eta^{\prime}$ \cite{BESIII:2020kpr} at different center-of-mass energies between 2.0 and 3.08 GeV, where an enhancement structure near 2034 MeV was observed in $e^+e^-\to \omega \pi^0$, and another enhancement structure near 2111 MeV 
 in $e^+e^-\to \eta^{\prime}\pi^+\pi^-$.
Both of these two processes are ideal processes to study $\rho$ meson states, which provide a good and precious opportunity to study the exited $\rho$ mesons around 2 GeV.
There are three $\rho$ meson states $\rho(1900)$, $\rho(2000)$, and $\rho(2150)$ included in PDG \cite{ParticleDataGroup:2020ssz} around 2 GeV, which are usually assigned as $\rho(3^3S_1)$, $\rho(2^3D_1)$, and $\rho(4^3S_1)$ based on mass spectrum studies, respectively.
However, as pointed out in the introduction, there are some problems in establishing the $\rho$ meson family around 2 GeV and understand these two structures observed in $e^+e^-\to \omega \pi^0$ \cite{BESIII:2020xmw} and $e^+e^-\to \rho\eta^{\prime}$ \cite{BESIII:2020kpr}.

In order to solve these problems, we have analyzed the experimental data of the Born cross sections of $e^+e^-\to \omega \pi^0$ and $e^+e^-\to \rho\eta^{\prime}$ with the theoretical support on mass spectrum, and decay behaviors of $\rho$ meson states around 2 GeV.
Our analysis has shown that the enhancement structure near 2034 MeV observed in $e^+e^-\to\omega\pi^0$ cannot be explained as a single state of $\rho(2000)$, which is dominated by the contributions of $\rho(1900)$ and $\rho(2150)$.
Another enhancement structure at 2111 MeV observed in $e^+e^-\to\rho\eta^{\prime}$ can be reproduced by the contributions from $\rho(2000)$ and $\rho(2150)$.
As a $D$-wave $\rho$ meson state, the $\rho(2000)$ decay to electron and positron should be suppressed, which makes it difficult to observe it in the $e^+e^-$ collision experiment.
However, our research has shown that the contribution of $\rho(2000)$ to $e^+e^-\to\rho\eta^{\prime}$ is comparable to that of $\rho(2150)$ because the $\rho(2000)$ has a strong coupling with the $\rho\eta^{\prime}$ channel.
Therefore, the $e^+e^- \to \rho\eta^{\prime}$ may be a golden process to look for the $\rho(2000)$.
On the other hand, we have found that the interference effect is very important to reproduce the experimental data, which may dramatically change the line shape of the Born cross sections.
The interference effect makes it difficult to accurately measure the resonance parameters in the experiments, especially when the statistics of experimental data are not abundant enough and multiple resonant states exist in a very small energy region.
This is one of the reasons why the resonance parameters of the $\rho$ meson states around 2 GeV are so poorly measured.

Establishing the $\rho$ meson states around 2 GeV is not only necessary for the study of $\rho$ meson states in the higher energy region, but also very helpful for the study of $\omega$ and $\phi$ meson states around 2 GeV.
The $e^+e^-\to\omega\pi^0$ and $e^+e^-\to\rho\eta^{\prime}$ are precious and clean two-body processes to study $\rho$ meson states.
We strongly urge researchers to focus on these two processes in the future, which could verify our analysis in this work, and more importantly, help establish the $\rho$ meson family around 2 GeV.

\section*{ACKNOWLEDGEMENTS}

This work is supported by the China National Funds for Distinguished Young Scientists under Grant No. 11825503, National Key Research and Development Program of China under Contract No. 2020YFA0406400, the 111 Project under Grant No. B20063, the National Natural Science Foundation of China under Grant No. 12047501, and by the Fundamental Research Funds for the Central Universities.


\begin{thebibliography}{199}

\bibitem{BESIII:2020xmw}
M.~Ablikim \textit{et al.} [BESIII],
Observation of a resonant structure in $e^{+}e^{-} \to \omega\eta$ and another in $e^{+}e^{-} \to \omega\pi^{0}$ at center-of-mass energies between 2.00 and 3.08 GeV,
Phys. Lett. B \textbf{813}, 136059 (2021).

\bibitem{BESIII:2020kpr}
M.~Ablikim \textit{et al.} [BESIII],
Measurement of the Born cross sections for $e^+e^- \to \eta^\prime \pi^{+}\pi^{-}$ at center-of-mass energies between $2.00$ and $3.08$\textasciitilde{}GeV,
Phys. Rev. D \textbf{103}, no.7, 072007 (2021).


\bibitem{ParticleDataGroup:2020ssz}
P.~A.~Zyla \textit{et al.} [Particle Data Group],
Review of Particle Physics,
PTEP \textbf{2020}, no.8, 083C01 (2020).




\bibitem{FENICE:1996xlc}
A.~Antonelli \textit{et al.} [FENICE],
Measurement of the total $e^+ e^- \to \rm{hadrons}$ cross-section near the $e^+ e^- \to N \bar{N}$ threshold,
Phys. Lett. B \textbf{365}, 427-430 (1996).

\bibitem{Frabetti:2001ah}
  P.~L.~Frabetti {\it et al.} [E687 Collaboration],
  Evidence for a Narrow Dip Structure at 1.9-GeV/$c^{2}$ in 3$\pi^{+} 3\pi^{-}$ Diffractive photoproduction,
  Phys.\ Lett.\ B {\bf 514}, 240 (2001).

\bibitem{Frabetti:2003pw}
  P.~L.~Frabetti {\it et al.},
  On the narrow dip structure at 1.9-GeV$/c^2$ in diffractive photoproduction,
  Phys.\ Lett.\ B {\bf 578}, 290 (2004).


\bibitem{BaBar:2007ceh}
B.~Aubert \textit{et al.} [BaBar],
Measurements of $e^{+} e^{-} \to K^{+} K^{-} \eta$, $K^{+} K^{-} \pi^0$ and $K^0_{s} K^\pm \pi^\mp$ cross- sections using initial state radiation events,
Phys. Rev. D \textbf{77}, 092002 (2008).

\bibitem{BaBar:2006vzy}
B.~Aubert \textit{et al.} [BaBar],
The $e^+e^- \to 3(\pi^+ \pi^-), 2(\pi^+ \pi^- \pi^0)$ and $K^+ K^- 2(\pi^+ \pi^-)$ cross sections at center-of-mass energies from production threshold to 4.5-GeV measured with initial-state radiation,
Phys. Rev. D \textbf{73}, 052003 (2006).


\bibitem{Hasan:1994he}
A.~Hasan and D.~V.~Bugg,
Amplitudes for $\bar{p} p \to \pi \pi$ from 0.36 GeV/c to 2.5 GeV/c,
Phys. Lett. B \textbf{334}, 215-219 (1994).


\bibitem{Bugg:2004xu}
D.~V.~Bugg,
Four sorts of meson,
Phys. Rept. \textbf{397}, 257-358 (2004).



\bibitem{He:2013ttg}
L.~P.~He, X.~Wang and X.~Liu,
Towards two-body strong decay behavior of higher $\rho$ and $\rho_3$ mesons,
Phys. Rev. D \textbf{88}, no.3, 034008 (2013).

\bibitem{Li:2021qgz}
Z.~Y.~Li, D.~M.~Li, E.~Wang, W.~C.~Yan and Q.~T.~Song,
Assignments of the Y(2040), \ensuremath{\rho}(1900), and \ensuremath{\rho}(2150) in the quark model,
Phys. Rev. D \textbf{104}, no.3, 034013 (2021).

\bibitem{Feng:2021igh}
J.~C.~Feng, X.~W.~Kang, Q.~F.~L\"u and F.~S.~Zhang,
Possible assignment of excited light S31 vector mesons,
Phys. Rev. D \textbf{104}, no.5, 054027 (2021).






\bibitem{Anisovich:2000kxa}
A.~V.~Anisovich, V.~V.~Anisovich and A.~V.~Sarantsev,
Systematics of $q \bar{q}$-states in the $(n, M^2)$ and $(J, M^2)$ planes,
Phys. Rev. D \textbf{62}, 051502 (2000).




\bibitem{Masjuan:2012gc}
P.~Masjuan, E.~Ruiz Arriola and W.~Broniowski,
Systematics of radial and angular-momentum Regge trajectories of light non-strange $q\bar{q}$-states*,
Phys. Rev. D \textbf{85}, 094006 (2012).

\bibitem{Masjuan:2013xta}
P.~Masjuan, E.~Ruiz Arriola and W.~Broniowski,
Reply to \textquotedblleft{}Comment on \textquoteleft{}Systematics of radial and angular-momentum Regge trajectories of light nonstrange $q\overline{q}$-states\textquoteright{} \textquotedblright{},
Phys. Rev. D \textbf{87}, no.11, 118502 (2013).


\bibitem{Bugg:2012yt}
D.~V.~Bugg,
Comment on \textquotedblleft{}Systematics of radial and angular-momentum Regge trajectories of light nonstrange $q\overline{q}$-states\textquotedblright{},
Phys. Rev. D \textbf{87}, no.11, 118501 (2013).


\bibitem{Wang:2021gle}
J.~Z.~Wang, L.~M.~Wang, X.~Liu and T.~Matsuki,
Deciphering the light vector meson contribution to the cross sections of e+e- annihilations into the open-strange channels through a combined analysis,
Phys. Rev. D \textbf{104}, no.5, 054045 (2021).



\bibitem{Yu:2021ggd}
G.~L.~Yu, Z.~G.~Wang, X.~W.~Wang and H.~J.~Wang,
The ground states and the first radially excited states of D-wave vector $\rho$ and $\phi$ mesons,
Int. J. Mod. Phys. A \textbf{36}, 2150197 (2021).


\bibitem{Wang:2020kte}
L.~M.~Wang, J.~Z.~Wang and X.~Liu,
Toward $e^+e^-\to \pi^+\pi^-$ annihilation inspired by higher $\rho$ mesonic states around 2.2 GeV,
Phys. Rev. D \textbf{102}, no.3, 034037 (2020).



\bibitem{Bauer:1975bv}
T.~Bauer and D.~R.~Yennie,
Corrections to VDM in the Photoproduction of Vector Mesons. 1. Mass Dependence of Amplitudes,
Phys. Lett. B \textbf{60}, 165-168 (1976).

\bibitem{Bauer:1975bw}
T.~Bauer and D.~R.~Yennie,
Corrections to diagonal VDM in the photoproduction of vector mesons. 2. Phi-omega Mixing,
Phys. Lett. B \textbf{60}, 169-171 (1976).

\bibitem{Kaymakcalan:1983qq}
O.~Kaymakcalan, S.~Rajeev and J.~Schechter,
Nonabelian Anomaly and Vector Meson Decays,
Phys. Rev. D \textbf{30}, 594 (1984).

\bibitem{Lin:1999ad}
Z.~w.~Lin and C.~M.~Ko,
A Model for $J / \psi$ absorption in hadronic matter,
Phys. Rev. C \textbf{62}, 034903 (2000).

\bibitem{Oh:2000qr}
Y.~s.~Oh, T.~Song and S.~H.~Lee,
$J / \psi$ absorption by pi and rho mesons in meson exchange model with anomalous parity interactions,
Phys. Rev. C \textbf{63}, 034901 (2001).




\bibitem{Chen:2011cj}
D.~Y.~Chen, X.~Liu and T.~Matsuki,
Two Charged Strangeonium-Like Structures Observable in the $Y(2175) \to \phi(1020)\pi^{+} \pi^{-}$ Process,
Eur. Phys. J. C \textbf{72}, 2008 (2012).



\bibitem{Godfrey:1985xj}
S.~Godfrey and N.~Isgur,
Mesons in a Relativized Quark Model with Chromodynamics,
Phys. Rev. D \textbf{32}, 189-231 (1985).



\bibitem{Achasov:2016zvn}
M.~N.~Achasov, A.~Y.~Barnyakov, K.~I.~Beloborodov, A.~V.~Berdyugin, D.~E.~Berkaev, A.~G.~Bogdanchikov, A.~A.~Botov, T.~V.~Dimova, V.~P.~Druzhinin and V.~B.~Golubev, \textit{et al.}
Updated measurement of the $e^+e^- \to \omega \pi^0 \to \pi^0\pi^0\gamma$ cross section with the SND detector,
Phys. Rev. D \textbf{94}, no.11, 112001 (2016).


\bibitem{BaBar:2017zmc}
J.~P.~Lees \textit{et al.} [BaBar],
Measurement of the ${e}^{+}{e}^{{-}}{\rightarrow}{{\pi}}^{+}{{\pi}}^{{-}}{{\pi}}^{0}{{\pi}}^{0}$ cross section using initial-state radiation at BABAR,
Phys. Rev. D \textbf{96}, no.9, 092009 (2017).



\bibitem{BaBar:2007qju}
B.~Aubert \textit{et al.} [BaBar],
The $e^+ e^- \to 2(\pi^+ \pi^-) \pi^0, 2(\pi^+ \pi^-) \eta, K^+ K^- \pi^+ \pi^- \pi^0$ and $K^+ K^- \pi^+ \pi^- \eta$ Cross Sections Measured with Initial-State Radiation,
Phys. Rev. D \textbf{76}, 092005 (2007),
[erratum: Phys. Rev. D \textbf{77}, 119902 (2008)].






\end{thebibliography}
\end{document}